# Gas-to-nanotextile: high-performance materials from floating 1D nanoparticles

Isabel Gómez-Palos,[a,b] Miguel Vazquez-Pufleau,[a] Richard S. Schäufele,[a,c] Anastasiia Mikhalchan,[a] Afshin Pendashteh,[a] Álvaro Ridruejo [b] and Juan J. Vilatela *[a]

[1] IMDEA Materials, Madrid, Spain

[2] Department of Materials Science, Universidad Politécnica de Madrid, E.T.S. de Ingenieros de Caminos, 28040 Madrid, Spain

[3] Department of Applied Physics, Universidad Autónoma de Madrid, Cantoblanco, Madrid, 28049 Spain

* *juanjose.vilatela@imdea.org*

Suspended in the gas phase, 1D inorganic nanoparticles (nanotubes and nanowires) grow to hundreds of microns in a second and can be thus directly assembled into freestanding network materials. The corresponding process continuously transforms gas precursors into aerosols into aerogels into macroscopic nanotextiles. By enabling the assembly of very high aspect ratio nanoparticles, this processing route has translated into high-performance structural materials, transparent conductors and battery anodes, amongst other embodiments. This paper reviews progress in the application of such manufacturing process to nanotubes and nanowires. It analyses 1D nanoparticle growth through floating catalyst chemical vapour deposition (FCCVD), in terms of reaction selectivity, scalability and its inherently ultra-fast growth rates ($10^7$–$10^8$ atoms per second) up to 1000 times faster than for substrate CVD. We summarise emerging descriptions of the formation of aerogels through percolation theory and multi-scale models for the collision and aggregation of 1D nanoparticles. The paper shows that macroscopic ensembles of 1D nanoparticles resemble textiles in their porous network structure, high flexibility and damage-tolerance. Their bulk properties depend strongly on inter-particle properties and are dominated by alignment and volume fraction. Selected examples of nanotextiles that surpass granular and monolithic materials include structural fibres with polymer-like toughness, transparent conductors, and slurry-free composite electrodes for energy storage.

# 1. Introduction

Nanotubes, nanowires and 2D nanosheets are highly crystalline inorganic building blocks with superlative axial (in-plane for 2D) properties, inherent defect tolerance, quantised electronic structure and extremely large aspect ratio. Many of their applications, such as electrodes, membranes, and structural elements, require forming macroscopic ensembles composed of a myriad of these building blocks. The challenge of assembling high aspect ratio particles into macroscopic materials resembles the early development of macromolecular materials based on synthetic polymers and gives a glimpse of the enormous transformative impact nanoparticles can have.

Processing or assembly of these high aspect ratio inorganic building blocks is challenging. They are much longer than traditional nanofillers, much thicker than polymer chains and much smaller than monolithic materials. Dispersion-based processes can be extremely effective at low volume fractions and have multiple applications, but can become increasingly ineffective as the nanoparticle length increases. One inherent limitation is the rapid increase in viscosity with increasing aspect ratio and/or volume fraction.[1] Similarly, thermodynamic considerations imply that higher aspect ratio nanoparticles require increasingly higher dilution to disperse them.[2] A typical dispersion of individualised high aspect ratio nanoparticles has a concentration of $10^{-2}$ mg ml$^{-1}$, or about $10^5$ times more volume of solvent than nanofillers, and requires previous stages of purification and centrifugation.[3]

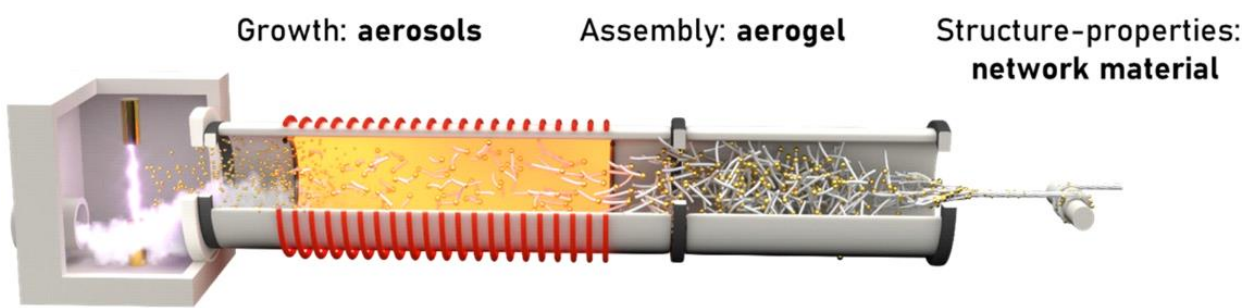

**Fig. 1** Schematic of a typical set-up for the synthesis of 1D nanoparticles based on the FCCVD method. The process begins with a catalyst aerosol generator, followed by a reaction zone where 1D nanoparticles grow, then comes a section where assembly occurs and finally, the synthesized material is obtained and tested for determining its structure and properties.

An emerging alternative is to synthesise 1D nanoparticles (nanotubes or nanowires) suspended in the gas phase and directly assemble them into macroscopic network materials, as shown schematically in Fig. 1, circumventing the need for dispersion processes. In essence, the process consists in the aerosol-catalysed synthesis of high aspect ratio nanoparticles that can aggregate to form an aerogel which can be shaped as a freestanding macroscopic nanotextile. Growth of 1D nanoparticles through floating catalyst chemical vapour deposition (FCCVD) is about 1000 times faster than substrate-based processes, leading to the synthesis of ultra-long nanoparticles that can form macroscopic networks and have translated into some of the best performing structural materials, transparent conductors and battery anodes. This manuscript sets out to review progress in using FCCVD for the synthesis of new high-performance macroscopic materials and identifies pressing open questions in the field. For clarity and to help unify similar concepts across different scientific communities, we include a glossary in the Appendix.

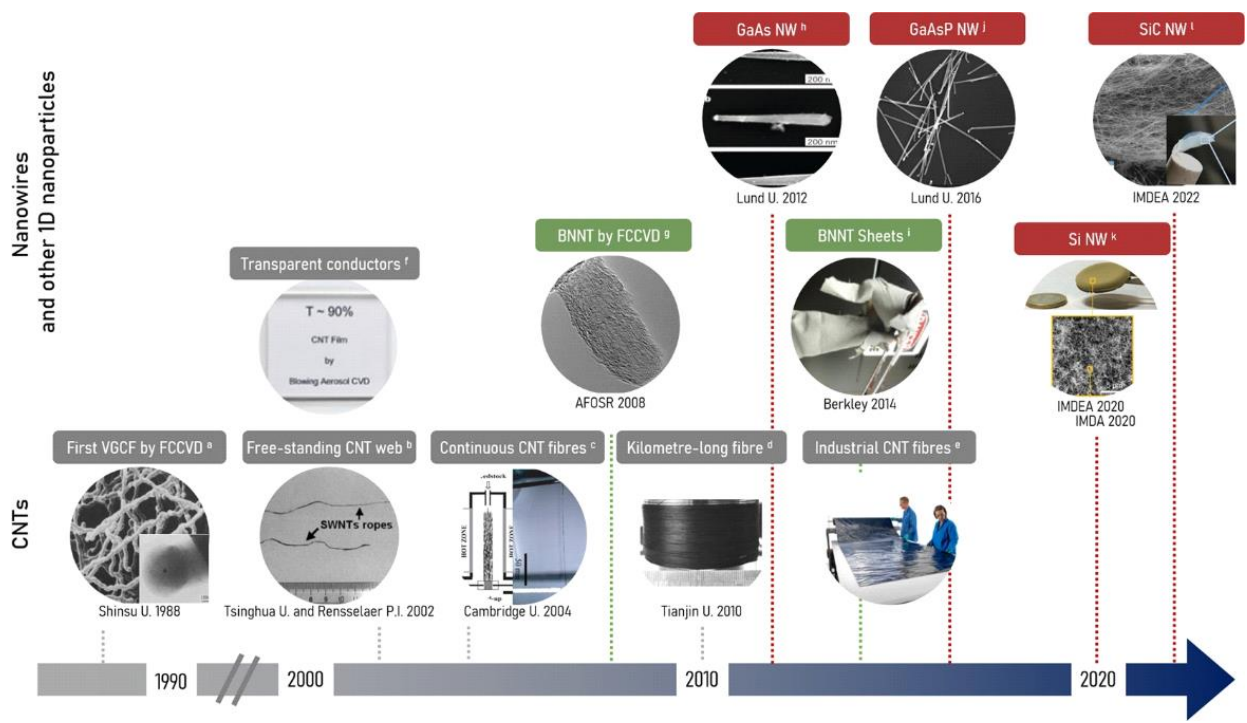

**Fig. 2** Timeline of the main developments in the synthesis of 1D nanoparticles by FCCVD and assembly into macroscopic materials. (A) VGCF.[5] (B) CNTs webs.[8] (C) CNTs continuous fibres.[9] (D) CNTs km-long fibre.[10] (E) CNTs industrial fibre. (F) CNTs transparent conductors.[11] (G) BNNTs.[12] (H) GaAs NW.[13] (I) BNNTs Sheets.[14] (J) GaAsP NW.[15] (K) Si NW.[16] (L) SiC NW.[17] (A) Adapted with permission from ref. 5. Copyright 1988 American Chemical Society. (B) Adapted with permission from ref. 8. Copyright 2002 American Association for the Advancement of Sciences. (C) Adapted with permission from ref. 9. Copyright 2004 American Association for the Advancement of Sciences. (D) Adapted with permission from ref. 10. Copyright 2010 Wiley-VCH. (E) Copyright 2023 Huntsman International LLC. (F) Adapted with permission from ref. 11. Copyright 2020 Wiley-VCH. (G) Adapted with permission from ref. 12. Copyright 2008 American Chemical Society. (H) Adapted with permission from ref. 13. Copyright 2012 Springer Nature. (I) Adapted with permission from ref. 14. Copyright 2014 American Chemical Society. (J) Adapted with permission from ref. 15. Copyright 2016 American Chemical Society. (K) Adapted from ref. 16 with permission from the Royal Society of Chemistry. (L) Adapted from ref. 17 with permission from the Royal Society of Chemistry.

## 2. Background

A timeline with breakthroughs in the synthesis and assembly of 1D nanoparticles by FCCVD is presented in Fig. 2. The initial evolution in synthesis of 1D nanoparticles by FCCVD occurred largely through parallel developments in the fields of nanocarbon and nanowire growth. The first reports of the synthesis of 1D nanoparticles via FCCVD date from the 80s, corresponding to vapour-grown carbon fibres (VGCFs) produced through the thermocatalytic decomposition of carbon sources in the presence of a transition metal aerosol in continuous flow reactors.[4,5] Early work on carbon nanotubes (CNTs) drew heavily from methods used for VGCF growth. Growth of CNTs from aerosol catalysts enabled large advances in the controlled synthesis of single-walled carbon nanotubes (SWCNTs)[6] and its early scale-up.[7]

A breakthrough came when the addition of sulphur during FCCVD produced a large increase in CNT length, and thus to their entanglement in the gas phase and formation of freestanding strands.[8] Soon after, use of different precursors and changes in synthesis reaction conditions led to the demonstration of spinning of continuous macroscopic sheets and fibres directly drawn from aerogels of CNTs synthesised by FCCVD.[9] Intense research activity on CNT fibres has followed ever since. Bulk fibre with better properties than the best engineering materials have been reported after improvements in CNT synthesis and assembly.[18,19] The CNT fibre manufacturing process continues to evolve but is now much better understood[20–22] and is used in scaled up industrial facilities. Improvements to control CNT molecular features (diameter and number of layers) and aggregation[23] under dilute reaction conditions have also translated into high-performance transparent conductors[24] and optoelectronic elements[25] based on networks of CNTs.

Stemming from CNT research, boron nitride nanotubes (BNNTs) were grown by FCCVD at Berkeley and the AFOSR.[12,26] Order-of-magnitude increases in growth rate were later obtained when using a high-pressure plasma reactor system.[14] Several studies have followed on the synthesis[27,28] of BNNTs and the properties of their macroscopic ensembles.[29,30]

In parallel to this work on nanotubes, Samuelson's research group at Lund University achieved the first demonstration of nanowire synthesis by FCCVD. Nanowires of GaAs were grown in the gas phase using an aerosol of Au nanoparticles as the catalyst; they coined the term "aerotaxy" to refer to this process.[13] Crucially, it was noted that growth rate in FCCVD was 20–1000 times higher than with substrate-bound catalysts. Extensive work followed on achieving compositional control,[15,31] understanding and controlling flow patterns in continuous reactors,[32] and scaling up manufacture for optoelectronic applications.[33–35]

Recently, we demonstrated the growth of two additional types of nanowires by FCCVD and their assembly into macroscopic structures directly from the gas phase, Si nanowires (SiNWs),[16] and separately, SiC nanowires (SiCNWs).[17] Their growth rate by FCCVD is also 2–3 orders of magnitude higher than in substrate-based CVD (SCVD). FCCVD-produced SiNW networks have extraordinary mechanical properties and extremely high electrochemical capacity for Li-ion storage as anodes in batteries.[36]

Evidence suggests the feasibility of synthesising over 100 different 1D inorganic nanoparticles *via* FCCVD. This implies that FCCVD can be a universal process for the synthesis and direct assembly of 1D nanoparticles into macroscopic network materials with high-performance properties. To review progress in this quest, we divide the process into three areas following the direction of the synthetic path (Fig. 1): first, the growth region where particles form an aerosol; next, the assembly of 1D nanoparticles to form aggregates and ultimately an aerogel; and finally the structure and bulk properties as denser network materials.

## 3. 1D nanoparticles grown from floating catalysts: the interface between CVD, catalysis and aerosol science

It is instructive to analyse 1D nanoparticle growth by FCCVD by comparison with conventional substrate-based CVD, its parent process. Based on current mechanistic understanding, in SCVD growth of 1D nanoparticles catalyst particles pre-deposited on a substrate are exposed to precursors at high temperature; the precursor material incorporates into the catalyst through migration of adatoms on the surface or through direct impingement; the excess of precursor material in the catalyst alloy leads to nucleation at the substrate interface and to epitaxial growth of sequential crystal monolayers. In FCCVD, the catalyst nanoparticles are suspended in the gas phase as an aerosol. The current view is that inclusion of adatoms comes mainly from direct influx. Nucleation occurs at the surface of the nanoparticles and the nanowire grows from progressive segregation of the precursor as a solid product. In both modes, the vapour–liquid–solid (VLS) growth mechanism is the most common. The differences between conventional substrate-based CVD and FCCVD are seemingly subtle, but have deep implications for the reaction conditions and the resulting solid products. Table 1 summarises the main features of the two synthesis modes.

**Table 1** Comparison of the main features of FCCVD and substrate CVD processes

|  | SCVD | FCCVD |
|---|---|---|
| **Reaction conditions** | | |
| Common growth mechanism | VLS | VLS |
| Reactor type | Close to CSTR | Plug flow reactor (PFR) |
| Residence time | ≈ 3000 s | ≈1-10 s |
| Pressure | high vacuum to atmospheric | close to atmospheric |
| Catalyst generation | Lithography, deposition on substrate | Aerosol |
| Reaction temperature (°C) | Limited by side reactions contaminating thin film | Allows higher Temp. |
| Growth velocity (nm s$^{-1}$) | 0.1 – 100 | 100 – 10$^6$ |
| Catalyst particle size | Control through patterning | Limited control through aerosol selection |
| **Reaction products** | | |
| Ensemble format | Vertically aligned and attached to substrate | From random to preferentially aligned networks |
| 1D nanoparticle materials synthesized | > 70 | CNT, GaAs(P)NW, BNNT, SiNW, SiCNW |

A fundamental difference is that FCCVD is conducted with the catalyst aerosol, precursors and reaction gases, all flowing through the reactor vessel (typically a ceramic tube). This reaction mode resembles a continuous tubular reactor typically idealised as a plug flow reactor (PFR). For FCCVD processes, it is employed using very short residence times and a non-constant temperature profile. In less than about 60 s, the species travel through the length of the tube and are exposed to a temperature profile with a heating gradient of up to 50 °C cm$^{-1}$. In that time, the precursors decompose, get incorporated in the catalyst and are segregated as ultra-long nanowires/ nanotubes. The same process typically takes 100 times longer in conventional SCVD. And very importantly, in SCVD there is usually continuous precursor feed to the stationary catalyst and the reaction zone temperature is nominally constant and uniform. This reaction mode resembles a continuous-stirred tank reactor (CSTR). Fig. 3 shows a schematic representation of the two reactor modes.

In substrate growth of nanowires at low pressure, growth rate follows a dependence of the form $r^{-p}$, where $r$ is the catalyst radius and $p$ is an exponent between 0.5 and 2 depending on growth conditions.[37] A known model suggests that this dependence is due to growth being limited by diffusion of adatoms to the catalyst from the substrate and the nanowire surface (Fig. 4A and B).[38] In such description, impingement of precursors or adatoms directly on the catalyst is explicitly neglected.

In FCCVD, the growth rate of GaAs,[32] SiNWs or SiCNWs at higher pressures shows no such dependence on nanowire radius (Fig. 4C). Johansson *et al.* have extended the substrate-based model to analyse the growth of inorganic nanowires from catalyst aerosols at high partial pressure.[39] Their analysis indicates that at the relatively high pressure used so far in FCCVD (>100 Pa) growth is limited by adatom incorporation, hence independent of pressure in this regime, as shown by the effect of partial pressure of group V precursors in Fig. 4D.

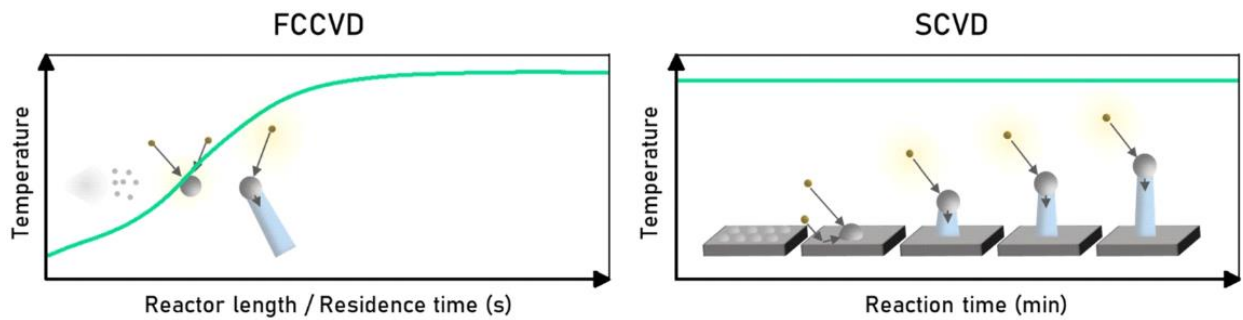

**Fig. 3** Scheme representing the growth mechanism by FCCVD and the more common substrate-based CVD.

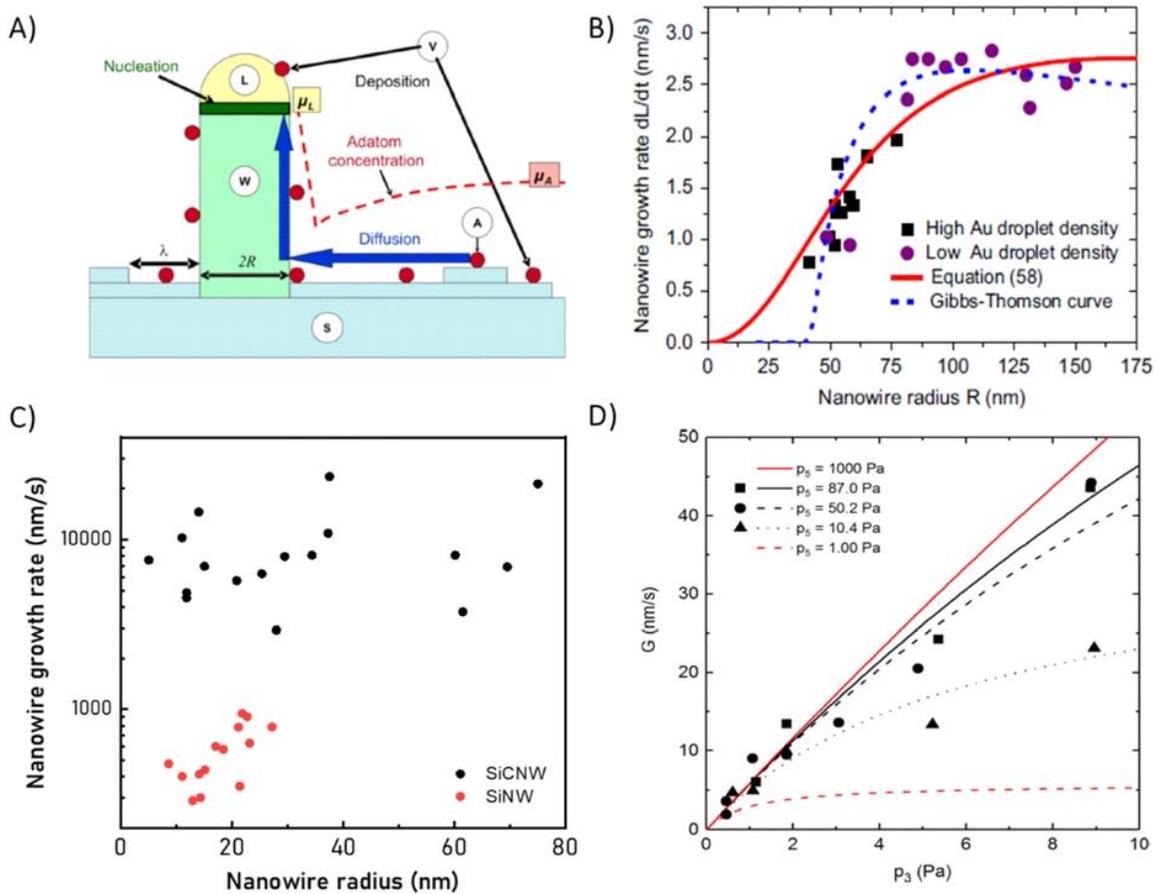

**Fig. 4** Dependence of nanowire growth rate for different VLS growth modes. (A and B) On substrates at low pressure, growth rate scales with r–p and can be limited by adatom diffusion to the catalyst.[37,38] (C) Nanowires of Si or SiC grown at higher pressure by FCCVD show no such dependence.[16,32] (D) Growth rate of GaAs nanowires via FCCVD as a function of partial pressure of group III and group V precursors. At high pressure in FCCVD, growth rate is limited by adatom incorporation and not by pressure.[39] (A) Reproduced with permission from ref. 38. Copyright 2008 American Physical Society. (B) Reproduced with permission from ref. 37. Copyright 2014 Elsevier. (D) Reproduced with permission from ref. 39. Copyright 2019 Elsevier.

Perhaps the most striking feature of FCCVD is the exceptionally fast growth rate observed for all nanowires and nanotubes so far produced with this method. This fast growth rate has also translated into high nanowire aspect ratios, which is desired for bulk applications, as discussed in Section 5.

Fig. 5 presents a comparison of reported growth rates for nanowires of GaAs, Si and SiC, and nanotubes of C and BN. For this comparison, we take the nanowire length increase over total growth time, thus assuming a constant growth rate. For FCCVD, given the exceptionally fast process, even determining an averaged growth rate constant is challenging. Lower limits of growth rate have been estimated by identifying the region in flow reactors where nanowires/nanotubes are formed, although with significant uncertainty.[17,20]

The comparison evidences that 1D nanoparticles inherently grow between 10 and 1000 times faster *via* FCCVD than SCVD. For VGCFs and CNTs, this has been previously attributed to the presence of sulphur[40] and to the higher reaction temperatures used in FCCVD, which could affect reaction kinetics or change the catalyst state from solid to liquid.[41] However, data for nanowire growth show a faster rate at equal reaction temperature, and at least for SiNWs, without suspected changes in solidification from the eutectic. In addition, all experimental data points in Fig. 5 correspond to growth conditions at high partial pressures. The indication is that this is a general feature of FCCVD. It produces order of magnitude faster growth rate than SCVD across all 1D nanoparticles, irrespective of the specific composition of the 1D crystal or the composition of the catalyst.

The two growth modes operate in different regimes. Overall, SCVD is considered to be in a regime limited by the diffusion of precursors and/or adatoms to the catalyst. In contrast, in FCCVD these diffusion limitations are not present, removing this rate-limiting step. Instead, the relatively narrow band of growth rates when plotted in atoms per second (in Fig. 5B) suggests that in FCCVD the limiting process is the rate of crystal step growth. This would mean that growth rate by FCCVD approaches the limit for VLS growth. The question, then, is about the differences in reaction conditions in FCCVD and SCVD leading to these two different regimes.

Early work proposed that FCCVD prevents the formation of a "depletion layer" by elimination of the substrate.[13] The premise is that in SCVD, the local concentration of the precursor is progressively reduced during nanowire growth, presumably due to the presence of precursor decomposition by-products in the vicinity of the catalyst and a longer diffusion path. This mechanism is plausible for a root-growth process, where the catalyst remains anchored to the substrate, but less so for tip-growth, which is also common for CNTs.

The concept of a depletion layer may not fully explain the differences in growth rate but does identify the importance of precursor availability. More precisely, FCCVD and SCVD are expected to have different rates of impingement of precursor molecules (or their derivatives) on the catalyst. In substrate-based CVD (in a CSTR), the temperature and catalyst concentration and size are constant, so the impingement rate is governed by precursor availability, which is constant under equilibrium in the reactor and assuming no changes in the diffusion path to reach the catalyst. For FCCVD, both the catalyst and the precursor are mobile and collide in the gas phase. The impingement rate is the collision rate. For typical distances between particles/molecules in FCCVD, such impingement rate might also depend on intermolecular forces between the catalyst and precursor.[57]

A way to rationalise the difference observed between FCCVD and SCVD processes is to observe the intrinsic differences between both processes. If one assumes that the critical step in nanowire growth is the formation of reactive radicals in the bulk gas, then the distance between such a radical and the closest available catalyst particle can be computed using Monte Carlo simulations for both FCCVD and SCVD processes using standard conditions for comparison. Fig. 6A shows the average catalyst-radical species separation assuming a concentration of $10^8$ part per $cm^3$, as reported in experimental FCCVD studies.[16,17] For the SCVD case, the assumption is a 1 $cm^3$ cube with 32 nm catalyst particles at the bottom with a density of 82 particles per $\mu m^2$. The predicted distance distribution of a randomly created radical due to Arrhenius type of activated collision is about 3 orders of magnitude smaller in the FCCVD case than in the SCVD case. Such a difference, along with the absence of major spatial constraints, *i.e.*, radicals can impinge from almost any direction, could be the reason why FCCVD has been experimentally shown to be also orders of magnitude faster than conventional SCVD processes. This distance difference would be critical in the case of non-ballistic type of catalyst–reagent radical interaction,[57] especially if a competing reaction can deplete the radical availability and reduce selectivity.[58] Another implication of shorter distances is the capability of maximizing precursor utilization with respect to the gas volume, resulting in higher production capacity.[13]

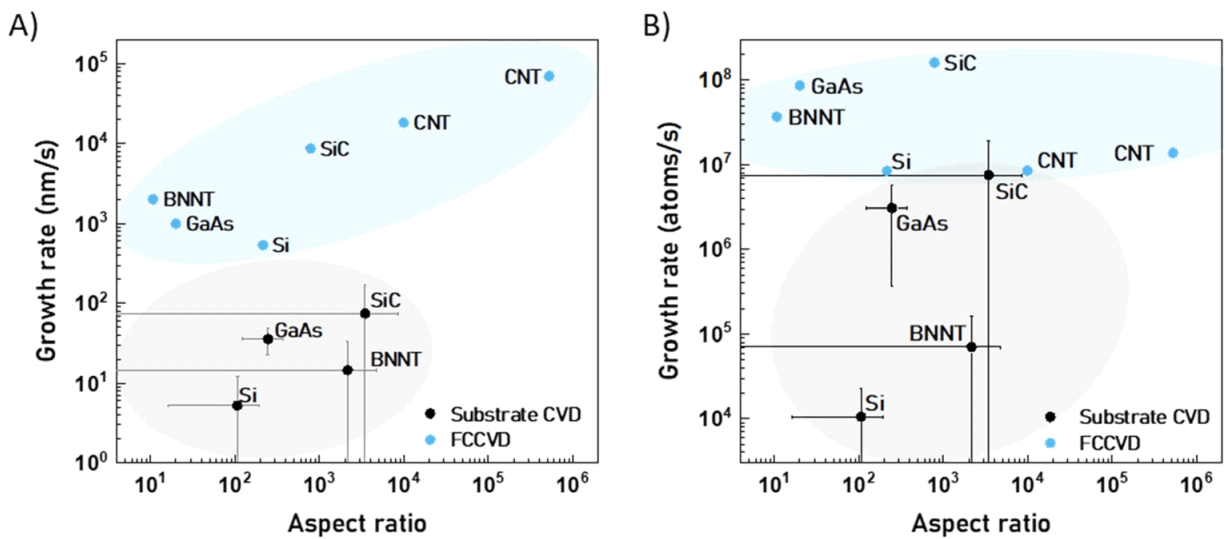

**Fig. 5**   Growth rate and aspect ratio of 1D nanoparticles of Si,[42] GaAs,[43,44] BNNT[45–50] and SiC[51–55] grown by substrate CVD (black) and Si,[16] GaAs,[13] BNNT,[12,26] CNTs,[24,56] and SiC[17] grown by FCCVD (blue). Using a catalyst floating in the gas phase produces order-of-magnitude faster growth. Note the growth rate axis units: (A) nm s−1 and (B) atoms per second.

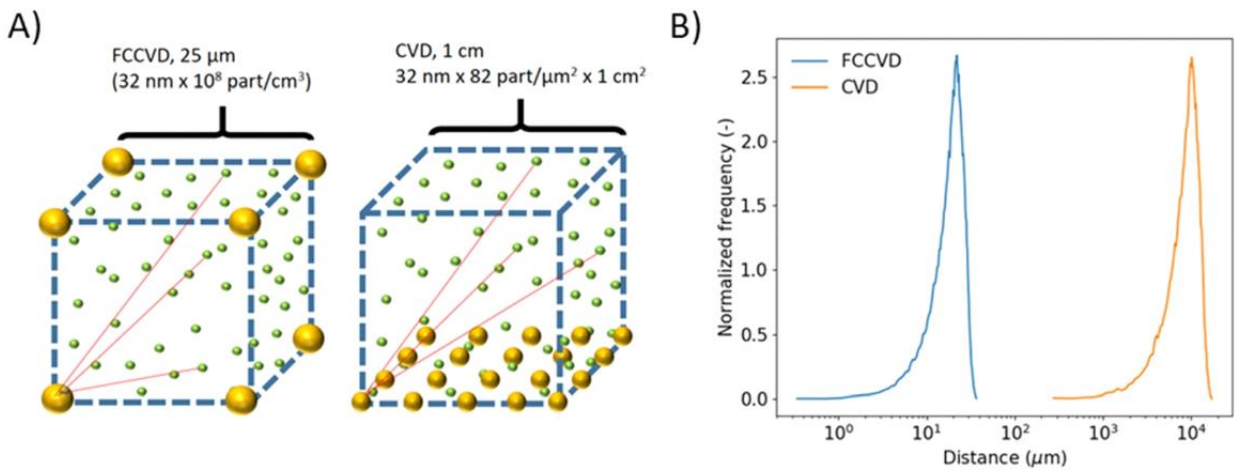

**Fig. 6**   Analysis of distance between gaseous radical species and catalysts. (A) Assumed systems for FCCVD and SCVD cases. (B) The distance distribution of a randomly generated reactive radical with respect to a catalyst particle for each case. The distance from FCCVD randomly generated radicals to catalyst particles is three orders of magnitude closer than by its counterpart by SCVD. For the FCCVD case a particle number concentration of 108 part per cm3 is used. For the SCVD case, the bottom is considered to be fully covered with catalyst particles (82 part per μm2 × 1 cm2). The same particle diameter is used in both cases and the distance distribution is then calculated using Monte Carlo generated radicals in the control volume.

## 3.1 Reaction selectivity

As in other transformation processes, FCCVD can produce other products in addition to 1D nanoparticles. The interest is in understanding the dominant factors enabling high reaction selectivity towards 1D nanoparticles. Of the multiple additional possible products, we focus on the predominant solid materials typically found in these reactions. The analysis ignores reactions at the walls of the reactor, which can be significant contributors to precursor depletion[57] and/or decomposition.[59]

The first type is non-catalysed solids, essentially nanoparticles that nucleate either in the gas-phase or as a layer on the surface of the nanoparticle. Various catalyst-free methods pursue the growth of similar nanoparticles as aerosols, but for the purpose of making macroscopic network materials, they are undesired impurities analogous to carbon soot. Examples of "soot" particles of GaAs, Si, and SiC found in the synthesis of different nanowires by FCCVD are included in Fig. 7.

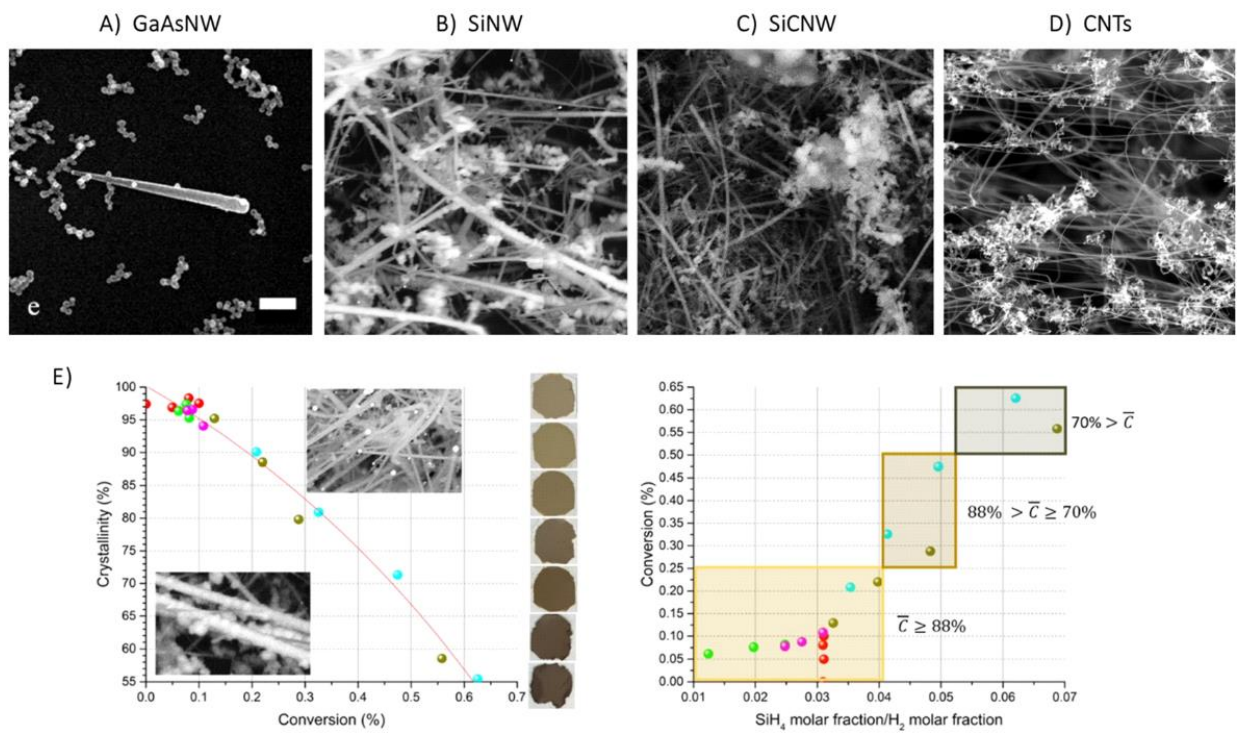

**Fig. 7**    Conversion and reaction selectivity in FCCVD growth of 1D nanoparticles. SEM examples of non-catalysed quasi-spherical nanoparticle byproducts, analogous to soot, formed in the synthesis of different 1D nanoparticles: (A) GaAsNW,[32] (B) SiNW,[58] (C) SiCNW, and (D) CNTs. (E) Plot of crystallinity, equivalent to the mass fraction of SiNWs (or 100% minus mass fraction of soot), and overall precursor to solid conversion for different precursor ($SiH_4$) and $H_2$ concentrations.[58] (A) Reproduced with permission from ref. 32. (B and E) Reproduced from ref. 58 with permission from the Royal Society of Chemistry.

The formation of soot over nanowires can be accurately analysed for Au-catalysed Si by FCCVD. At the low temperature required for SiNW growth the Si soot formed is amorphous, hence enabling discrimination of the two products and quantification of the mass fraction of each product, for instance using Raman spectroscopy and XRD. (For the study in ref. 58, the mass fraction of SiNWs relative to all solid products is equal to the degree of sample crystallinity, $\bar{C}$.) Therefore, these measurements provide a direct metric for selectivity under different reaction conditions.

Soot and related solid products are preferentially formed in the absence of hydrogen, hence why a hydrogen atmosphere is used in FCCVD growth of CNTs,[20] SiCNWs[17] and SiNWs[58] and gas-phase growth of BNNTs.[60] They also form at high precursor concentrations (in nitrogen) for GaAsNW FCCVD synthesis.[32]

In SiNW growth by FCCVD, at a low $SiH_4/H_2$ ratio, nearly 100% selectivity to SiNWs can be achieved. Higher ratios, for instance, by augmenting precursor availability, produce increasingly more nucleation of amorphous nanoparticles, *i.e.*, Si soot, which grows in addition to the SiNWs (Fig. 7E). The quantification of catalysed and non-catalysed solid products, combined with a relatively simple decomposition route for $SiH_4$, enables a quantitative description of the role of hydrogen in controlling reaction selectivity for SiNWs. Hydrogen inhibits the decomposition of silane into derivatives that would otherwise polymerise into amorphous Si nanoparticles, according to the reactions in Fig. 8A. The experimental data for conversion and selectivity could be fitted to a reaction model using a combination of literature data for the kinetics of non-catalysed formation of Si nanoparticles, an expression for growth velocity of nanowires and a proposed reaction mechanism based on the silylene radical (Fig. 8B).[58]

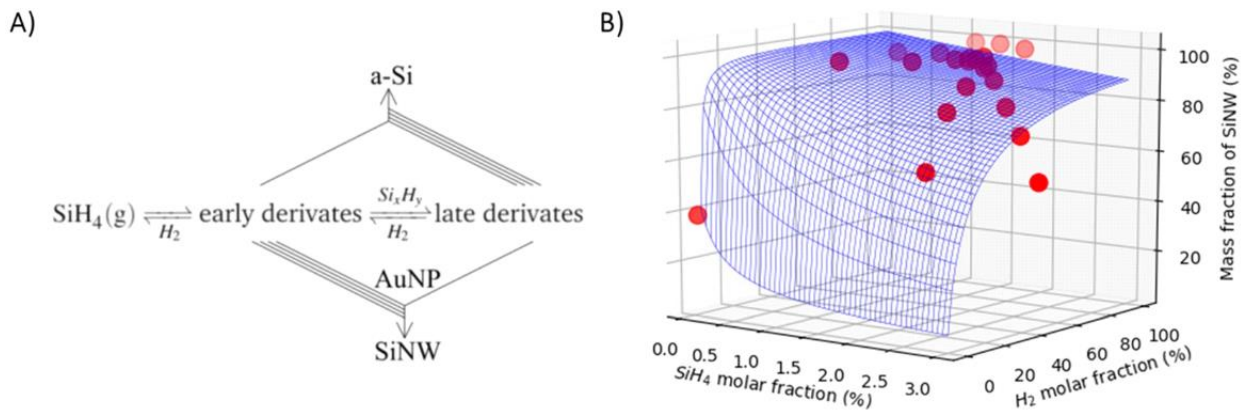

**Fig. 8** Competing formation of solid products in SiNW synthesis by FCCVD. (A) Proposed reaction scheme for Au-catalysed growth of SiNWs and non-catalyst formation of amorphous Si nanoparticles. (B) Experimental data of reaction selectivity, taken as crystallinity = SiNW mass/(amorphous + SiNW mass), and fitting with a kinetic model. Reproduced from ref. 58 with permission from the Royal Society of Chemistry.

Looking ahead, it is of interest to produce a similar description of the competing reactions in the growth of other 1D nanoparticles. Amongst the data required from experiment and modelling are (i) a measure of the mass and particle number concentration of solid reaction products, (ii) kinetic expressions describing the rate of soot formation, (iii) kinetic expressions for nanowire growth, and (iv) a tentative reaction mechanism tree with dominant paths.

The other type of undesired products is solids produced through the catalyst, but different from the target nanowires. These may consist of a nucleated shell or nanowires of different morphology or composition from the desired one.

In the growth of nanowires by FCCVD, the observation is that under successful conditions for nanowire growth, all the catalyst that travels through the flow reactor produces nanowires. As shown in the example in Fig. 9, inspection of solid products collected from the gas phase consistently shows that all catalyst nanoparticles are at the tip of nanowires. This is generally observed for growth of 1D nanoparticles of Si, SiC and GaAs. In contrast, in FCCVD growth of carbon nanotubes, the vast majority of the catalyst does not produce CNTs. The example in Fig. 9 shows a typical aerogel of CNTs with multiple residual catalyst particles attached to them. From the knowledge of CNT dimensions and the mass fraction of C and Fe, it is estimated that only 0.05 at% of Fe nanoparticles produce CNTs.[56] From *ex situ* analysis, most of the catalyst corresponds to metallic nanoparticles encapsulated with a graphitic shell (Fig. 9B). Residual core–shell particles have been found irrespective of the choice of transition metals used as the catalyst in FCCVD growth of CNTs.

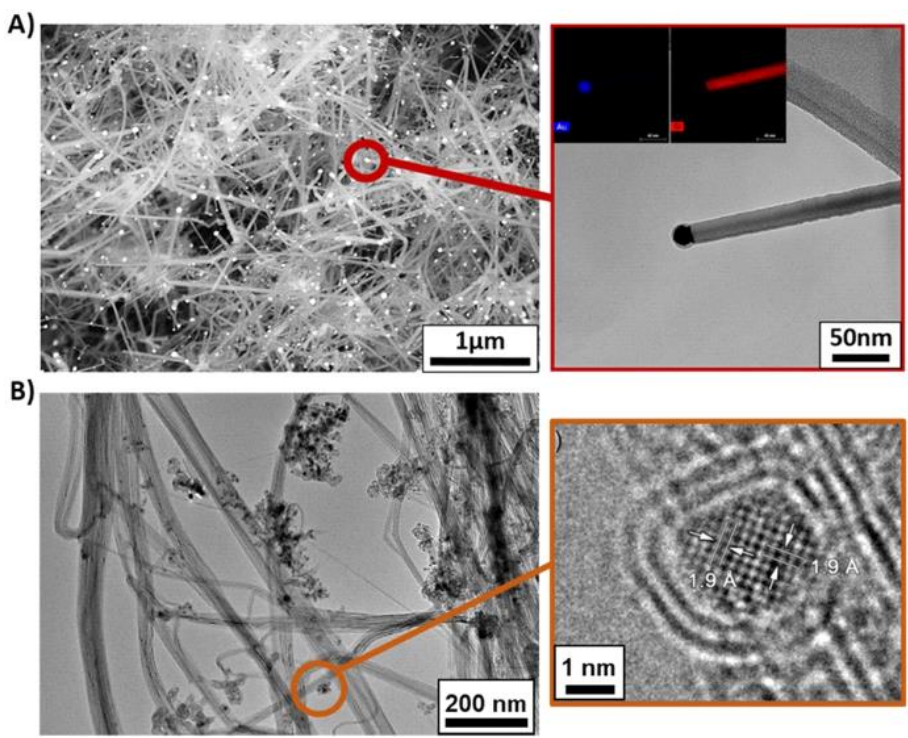

**Fig. 9** Examples of electron micrographs showing the high fraction of active catalyst (≈100%) in nanowire growth and extremely low fraction (<0.1%) of active catalyst in nanotube synthesis, both grown by FCCVD. SEM and TEM micrographs for (A) silicon nanowires and (B) CNTs and inactive catalyst with a carbon shell. (A) Reproduced from ref. 58 with permission from the Royal Society of Chemistry. (B) Adapted with permission from ref. 56. Copyright 2014 American Chemical Society.

Given their extreme abundance, we consider possible formation mechanisms as alternative reaction paths to CNT growth. One possibility is that the shell segregates upon cooling. For such a case, these catalysts may be considered inactive for the production of any solid nucleus, let alone growing CNTs. There are no *in situ* experimental measurements to discard this scenario, but it seems the more improbable. From a simple mass balance calculation, the minimum mass fraction of C in a metal catalyst nanoparticle to produce a shell is approximately $1/(1 + 2\Phi)$ (with $\Phi$, the nanoparticle diameter, in nanometres).[61] For small particles, for instance $\Phi = 1$ nm, the mass fraction (33%) is much larger than the concentration in the Fe–C eutectic (4.3), although they become comparable for particle diameters above around 8 nm.

More likely, the shell forms in the reaction zone, causing encapsulation of the catalyst and thus impeding CNT growth. After all, an encapsulating graphitic shell can be the most likely solid product for an sp²-hybridised crystal in the absence of an interface with a substrate. As a corollary, it follows that this alternative reaction path is governed by the crystal structure and surface energy of the nucleus and 1D nanostructure, hence prominent for nanotubes but not for nanowires. From this perspective, selectivity towards nanotubes requires preventing catalyst encapsulation. Promotors used in CNT growth by FCCVD are thought to form biphasic catalyst particles,[21,62] which may reduce encapsulation probability. Indeed, the addition of sulphur as the promoter increases conversion in FCCVD CNT growth;[63] however, it is challenging to separate its effect on CNT thickness and number concentration.

The bulk equilibrium phase diagram provides more insight into the reaction conditions leading to selective growth of specific nanowires. Analysing the nanowire nucleation process as a solidification of a bulk alloy is a simplification, but some experimental observations argue in favour of it. For example, the minimum experimental temperatures used for growth of various nanotubes and nanowires by FCCVD are close to the eutectic temperature on the corresponding phase diagram (Table 2). In addition, decomposition of precursors is likely to be fast compared to the residence times. Decomposition of ferrocene, a source of Fe catalyst, is estimated at a maximum of 0.1 s under experimental conditions,[20] for example.

**Table 2** Experimental growth temperatures in FCCVD and corresponding temperatures of eutectics

| 1D nanoparticle | Catalyst | Experimental growth temperature | Eutectic temperature from phase diagram |
| --- | --- | --- | --- |
| **CNT** | Fe | 1050–1250 °C | 1147 °C (Fe–C) |
| **BNNT** | Fe | 1200 °C | 1018 °C (Ni–B) |
| **SiNW** | Au | 650 °C | 360 °C (Au–Si) |
| **SiCNW** | Fe | 1130 °C | 1170 °C (Fe–SiC) |
| **GaAs** | Au | 425–625 °C | 350 & 450 °C (Au–Ga) |

One strategy to use thermodynamic equilibrium phase data is to represent the nanowire nucleation process as a trajectory in the isothermal phase diagram starting at the catalyst end and moving towards the constituent(s) of the nanowire. Fig. 10 shows examples for CNTs and SiCNWs. Relevant for CNTs grown by FCCVD, the Fe–S–C ternary diagram at high temperature (Fig. 10A) presents an Fe-rich liquid (L), and two immiscible liquids, one rich in C ($L_1$) and the other in S ($L_2$). At high C contents, the immiscible liquids are in equilibrium with solid C ($L_1 + L_2 + C_{(s)}$). The hypothesis indicates that this composition range produces a graphitic shell, which thus encapsulates the catalyst and prevents CNT growth. Growth of CNTs is thought to occur for lower S compositions at the boundary between L and $L_1 + L_2$, where solid C is not in equilibrium and thus gets ejected as the sidewalls of a CNT. The envisaged mechanism for CNT growth in these conditions is shown in Fig. 10B. An S-containing shell is predicted by the phase diagram upon cooling and observed experimentally.[62,64]

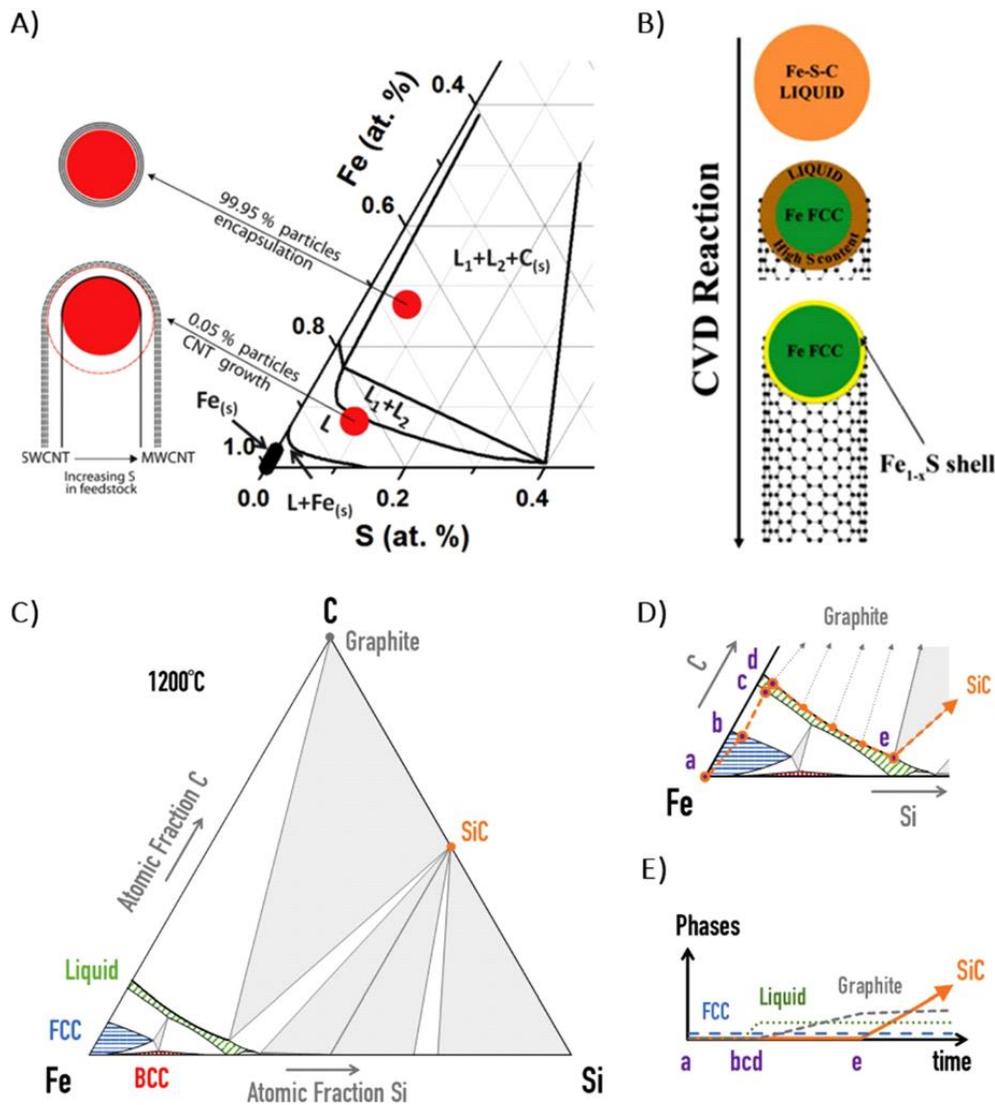

**Fig. 10** Thermodynamics of bulk systems and their relation to preferential nucleation of CNTs and SiCNWs. (A) Ternary bulk Fe–S–C phase diagram and (B) schematic of the corresponding growth process for CNTs. (C) Ternary bulk Fe–Si–C phase diagram with the estimated reaction paths marked enlarged in (D and E) predicted phase fraction evolution on cooling. (A) Reproduced with permission from ref. 56. Copyright 2014 American Chemical Society. (B) Reproduced from ref. 64 with permission from the Royal Society of Chemistry. (C–E) Reproduced from ref. 17 with permission from the Royal Society of Chemistry.

The calculated Fe–Si–C phase diagram has also been recently applied to rationalise the growth of SiCNWs (Fig. 10C– E).[17] The SiC nanowires were produced using an aerosol of Fe using a precursor containing both C and Si, under a hydrogen atmosphere. The synthesis conditions used did not produce CNTs or SiNWs, despite being close to those used when targeting such 1D nanoparticles. Fig. 10D shows the calculated trajectory on the Fe–Si–C phase diagram starting from the catalyst and increasing Si–C content according to the stoichiometry of the precursor. The phase fractions are illustrated in Fig. 10E. For this and other trajectories, a liquid is first formed. Upon segregation of C the liquid moves down to the Si-rich corner. When the Si solubility of the liquid is exceeded, SiC is then segregated. The implication is that provided both Si and C are incorporated into the catalyst, the expected solid product is predominantly SiC.

By way of summary, in Fig. 11 we present a schematic of the main solid products and corresponding reaction paths for nanotubes and nanowires. It includes estimates of the catalyst mass fraction distributions, which highlight how seemingly similar growth processes, namely nanotube and nanowire growth, lead to diametrically opposite use of catalysts supplied and different competing reactions that can reduce selectivity.

## 3.2 Nanowire/nanotube control

The most direct way to control the diameter of 1D nanoparticles produced by VLS is through the size of the catalyst particle. For substrate-based CVD growth, seed particles are deposited/formed onto the substrate prior to growth, from colloidal dispersions, thin films or using lithography. This enables large control over particle size distribution and position on the substrate. For the floating catalyst, the nanoparticles are formed as an aerosol either *via* chemical means such as *in situ* decomposition of catalyst precursors (*e.g.*, ferrocene),[21] or by physical means, including the formation of an aerosol through thermal evaporation[16] or spark discharge[65] of the catalyst material before entry into the FCCVD reactor.[66,67] These aerosol methods produce catalyst particles with a broad size distribution. The size distribution may be reduced using particle size selection methods or in-line sintering, but at the expense of reducing catalyst concentration. Typical particle size distributions are shown in Fig. 12A.

In nanowire growth by FCCVD, there is a broad correspondence between nanowire diameter and catalyst size, provided NW tapering (the process by which the bottom of the NW is thicker than the top) is avoided. This correspondence has been observed for Si, SiC and GaAs. An example of a SiNW diameter distribution is included in Fig. 12B. The nanowire diameter distributions are log–normal, with geometric standard deviation > 1.36, as summarized in Table 3, which follows the expectation for an aerosol of metallic catalyst particles.[68]

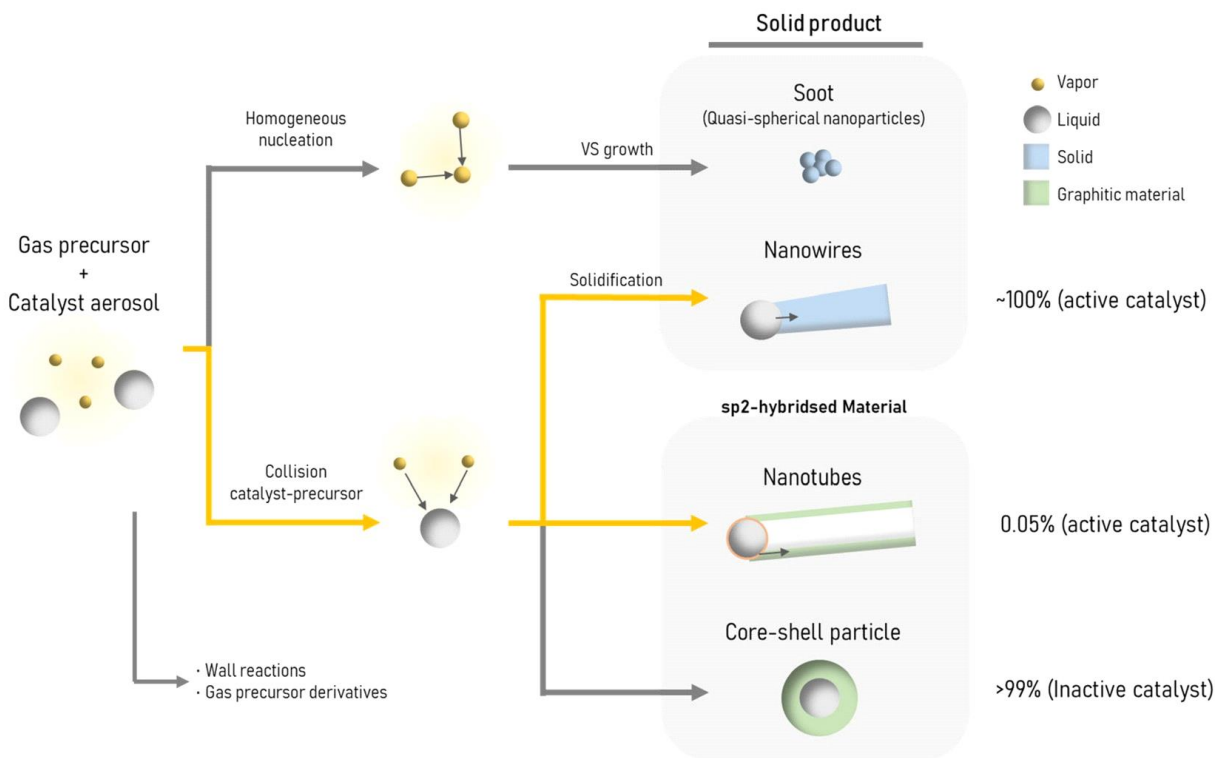

**Fig. 11** Schematic representation of the factors affecting selectivity in FCCVD growth of nanowires and nanotubes.

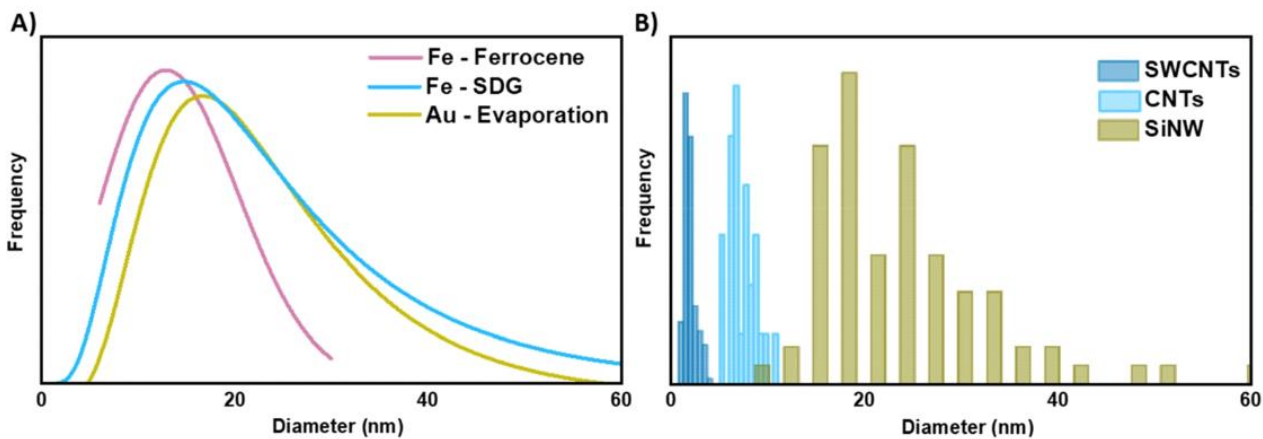

**Fig. 12** Comparison of catalyst size and nanowire/nanotube diameter distributions. (A) Comparison of catalyst size distributions achievable with different generation methods (*e.g.*, ferrocene decomposition,[69] spark discharge generation (SDG) and thermal evaporation); (B) diameter distribution of nanowires (Si) and nanotubes (CNT,[70] SWCNTs[71]) produced by FCCVD from an aerosol of catalysts without size selection and therefore a broad distribution. Note the correspondence between nanowire diameter and catalyst size, but not for nanotube diameter.

**Table 3** Mean diameter and geometric standard deviation (GSD) of log–normal distributions of 1D nanoparticles produced by FCCVD

| 1D nanoparticle | Mean diameter (nm) | Geometric standard deviation (GSD) |
| --- | --- | --- |
| Si NW | 20–25 | 1.33–1.41 |
| GaAs NW | 65 | |
| SiC NW | 17–45 | 1.43–1.78 |
| CNTs | 1.9–7.1 | 1.21–1.36 |
| BNNT | 2.0 | |

For nanotube growth by FCCVD, its outer diameter is approximately the same as the size of its catalyst particle. However, as noted before, only a very small fraction of aerosol particles form nanotubes. This is also evident when considering the diameter distribution of nanotubes produced by FCCVD. As shown in Fig. 12B, this distribution is at the low end compared to that of nanowires and represents a minute fraction of the catalyst aerosol in the absence of size selection methods.

Whereas in nanowire growth by FCCVD diameter can be controlled using established aerosol methods,[35] controlling inner and outer diameters of nanotubes is more challenging and additionally requires intervention at the stage of incorporation of the precursor and the composition of the catalyst.

This has been achieved through control of injection of precursors and gases,[22] optimization of residence times,[72] and introduction of group 16 elements.[21]

At present, the FCCVD process would seem to offer limited scope to narrow the distribution of 1D nanoparticle diameter closer to a monodisperse distribution, which is a requirement for some applications in optoelectronics. For those applications, SCVD with pre-formed monodispersed catalysts prevented from coalescence represents a better alternative.

## 3.3 Throughput

As a general chemical process, CVD-based growth of 1D nanoparticles is considered scalable. CNTs are produced in fluidised bed reactors in fully industrialised facilities with a capacity of the order of 2 kt per year per plant. Synthesis of SiNWs in laboratory-scale fluidised bed reactors with a scalable process based on silica nanoparticles resembling geodes has been recently demonstrated.[73] Specifically concerning FCCVD, the oldest scaled-up example is VGCF production. Industrial-grade VGCFs have been available for decades at relatively large industrial facilities with a reported capacity of tens of tons per site as early as 2001.[74]

Despite the seemingly inherent scalability of CVD-based reactions, FCCVD growth of 1D nanoparticles is mostly operated at a low throughput in research facilities. Throughput in laboratory-scale reactors is at grams per day and at several pilot plant industrial facilities still estimated below a kilogram per day. A detailed analysis of the scalability of FCCVD is beyond the scope of this paper, but a review of reactor operation conditions on the laboratory scale can provide general indicators of progress independent of reactor size.

A figure of merit for productivity of industrial catalytic processes is the ratio of mass produced per reactor volume and unit time. Analysis of industrial production of fuels and chemicals suggests a threshold for industrial viability of 0.1 t $(m^3\ h)^{-1}$.[75] Laboratory-scale reactors for FCCVD synthesis of CNTs and NWs typically produce less than 1 g $h^{-1}$, with a productivity of around $10^{-5}$ t $(m^3\ h)^{-1}$ (Table 4).

Table 4  Comparison of relevant values of 1D nanomaterial synthesis vs. carbon black and $TiO_2$ industrial synthesis

| Material | Throughput (g $h^{-1}$) | Productivity (t $(m^3\ h)^{-1}$) | Residence time (s) | Process yield (%) |
|---|---|---|---|---|
| VGCF[a] | 500 | $8.0 \times 10^{-3}$ | 29 | 31.3 |
| CNT | 0.06 | $5.2 \times 10^{-5}$ | 53 | 3.2 |
| SiNW | 0.2 | $8.5 \times 10^{-5}$ | 10 | 4.1 |
| BNNT[26] | 0.02 | $3.4 \times 10^{-5}$ | 167 | 0.5 |
| Carbon black[76] [b] | $8 \times 10^6$ | | 0.01–0.9 | 20–65 |
| Industrial $TiO_2$[77] [b] | $9 \times 10^6$ | | 0.160 | |

[a] Data for pilot plant reactor, courtesy of Grupo Antolín. [b] European Commission – European IPPC Bureau – large volume inorganic chemicals, solids and other industry.

Common values for the parameters currently employed in the process, as well as others related to precursor conversion for nanoparticle growth, are included in Table 4 and compared to industrial processes (note that the residence time included in Table 4 corresponds to the estimated reaction residence time, rather than the total time of precursor flow in the reactor).

Under perfect conversion and 100% selectivity, the process may be limited by the rate of solidification, in which case the throughput is the product of the number of catalyst particles and nanowire growth rate. From the experimental data in Fig. 5, the nanowire growth rate is of the order of $10^{-13}$ g $s^{-1}$. For the residence times used in FCCVD of 1–10 s and assuming no losses to the walls, the catalyst concentration can reach $10^9$ $cm^{-3}$. Taking the conventional tubular furnace reactor dimensions and flow rates, the mass flux comes out at 1.8 kg $h^{-1}$, which gives a productivity of 0.11 t $(m^3\ h)^{-1}$, just above the industrial threshold for fuels and chemicals. These may be considered upper limits for laboratory-scale reactors. Indeed, small pilot plant reactors for VGCFs with a throughput of 500 g h$^{-1}$ have productivities of 10$^{-3}$ t (m$^3$ h)$^{-1}$, although presumably higher at full scale up.

For some FCCVD reactions, an opportunity to increase productivity is by increasing the fraction of catalyst particles that produce nanowires ($f$). As noted before, current FCCVD growth of CNTs (and BNNT) operates under values of $f \sim 0.001$ due to low selectivity. For nanowire growth by FCCVD, virtually all catalyst particles that remain as an aerosol produce nanowires and $f \approx 1$. However, their productivity is at present comparable because of the $\approx 10^3$ times higher aspect ratio of CNTs.

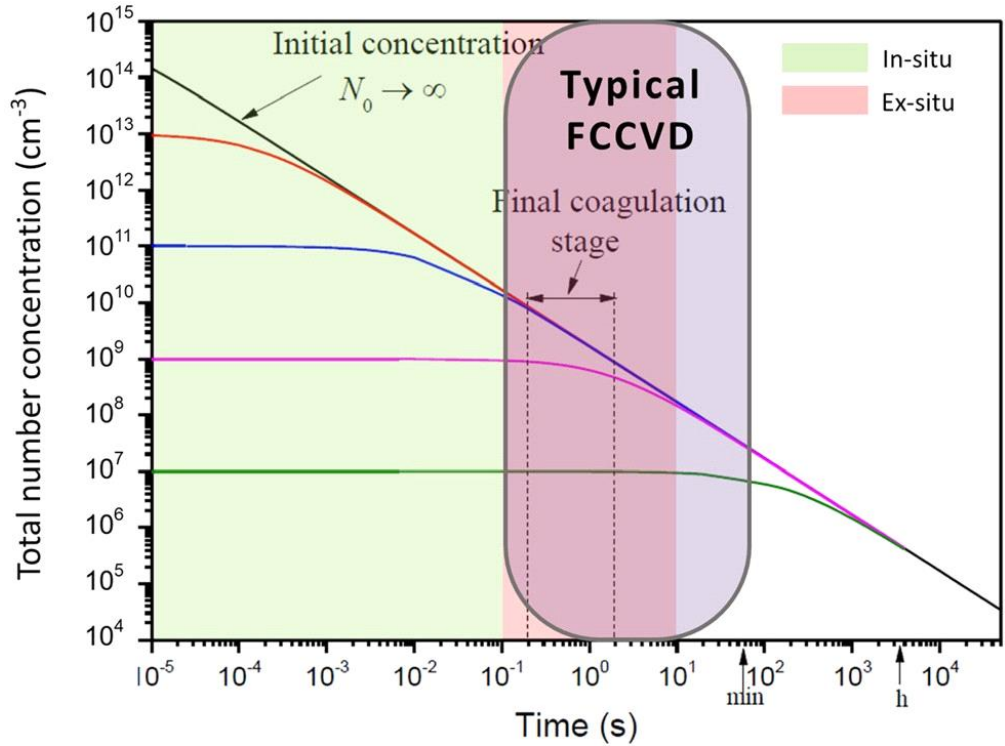

**Fig. 13**  Total number concentration of the floating catalyst (aerosol) that can exist in a system as a function of time for different initial number concentrations. The shaded oval is the theoretical limit for the number concentration of catalyst nanoparticles and thus of catalysed nanowires, for the corresponding residence times. Adapted with permission from ref. 78.

Higher productivity could be achieved with a shorter residence time. Industrial flame pyrolysis processes, for example, require 10$^4$ shorter residence times. However, it is possible that residence times of 1–10 s are close to the limits imposed by the rate of crystal growth. Note, for instance, that although residence times for spherical nanoparticles are 10$^4$ shorter than for nanotubes/nanowires, their aspect ratio is smaller by a similar amount. This may indicate that residence times in both processes are limited by solidification rates.

Under conditions when there is no excess of catalyst, for instance, in growth of NWs, smaller catalyst particles may lead to higher productivity on account of their higher surface-to-volume ratio. There are several aerosol methods for control of particle size; however, preserving the catalyst aerosol size distribution through the reactor is challenging considering the high target particle concentrations and the expected particle coalescence upon coagulation given the high operating temperature in FCCVD. The catalyst aerosol produced will evolve and converge into a single line, as depicted in Fig. 13, with residence times used in FCCVD of 1–10 s corresponding to a maximum catalyst concentration of $\approx 10^8$–$10^9$ cm$^{-3}$. A review of the literature shows that the catalyst concentration used in FCCVD growth is $10^6$ cm$^{-3}$ for GaAs,[13] $10^7$ cm$^{-3}$ for SiNW[16] and $10^6$–$10^9$ cm$^{-3}$ for CNT, [66,69] suggesting that much higher catalyst concentrations than those currently used are plausible.

Often for FCCVD synthesis, productivity is limited by precursor conversion. Present FCCVD growth processes are operated under conditions where the precursor concentrations are extremely low. As identified for CNTs, besides increases in conversion and collection, reaching significant throughput will require large increases in precursor and catalyst concentration in the reaction zone.[20] This is evident by considering both the process efficiency and the process yield. The former is the fraction of the precursor transformed into the desired solid, and the latter is the fraction of solid products over all species in the reaction zone. Their ratio is the molar fraction of precursors. Fig. 14 shows experimental data for process yield and precursor efficiency for SiNWs and CNTs. Interestingly, the data fall on a straight line, which may be partly due to experiments on SiNW growth being conducted using parameters taken from CNT growth as reference. Overall, the data show that less than 10% of the precursor is transformed into collected solid and that less than 1% of the species introduced form nanowires/nanotubes. Recycling of carrier gases, for example, will increase process yield and move the data upwards. Increasing conversion or collection of nanowires improves efficiency along the diagonal.

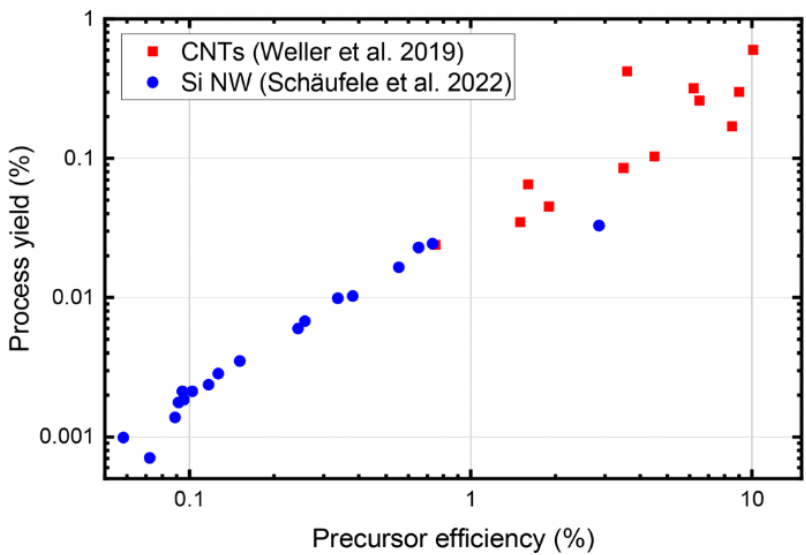

**Fig. 14**  Process yield and precursor efficiency in FCCVD synthesis of 1D nanoparticles.[20,58]

In view of this analysis, the challenge to increase productivity for 1D nanoparticle growth by FCCVD may be summarised as achieving >10% precursor conversion with near 100% selectivity at catalyst concentrations above $10^8$ cm$^{-3}$.

## 4. Assembly from the gas-phase: from aerosols to aerogels

Boies *et al.* have recently performed a seminal kinetic analysis of aerogel formation based on determination of collision rates of 1D nanoparticles.[79] They used simulation to determine the first collision kernel for rigid 1D nanoparticles, accounting for both particle rotation[80] and translation, for different 1D morphologies. Reduced order relations were provided to enable direct calculation of collision kernels for 1D nanoparticles from geometric parameters without the need for simulations.

Their analysis shows that 1D nanoparticles commonly produced by FCCVD fall in an intermediate regime, where the timescales for rotation ($\hat{t}_r$) and translation ($\hat{t}_x$) are comparable. As shown in Fig. 15A, $\hat{t}_r/\hat{t}_x \sim 0.3$ for the common diameters and lengths of nanotubes and nanowires. Fig. 15B shows the dimensional collision kernel for rigid 1D nanoparticles of different length. It shows that collisions are enhanced for nanoparticles of dissimilar length. In addition, the resulting kernel gives a 10-fold enhancement in collisions compared to spheres (dashed lines). The authors also analysed the formation of bundles: aligned aggregates of 1D nanoparticles that enhance particle interaction and are commonly found in CNTs. With the knowledge of the collision kernel, the authors calculated the time for bundling (Fig. 15C) for the conditions found in FCCVD reactors with a concentration of $10^8$ cm$^{-3}$ and accounting for a range of morphological parameters. The mean bundling time obtained of 3 seconds agrees with the residence times found in such reactors (Fig. 15D).

In a follow-up study the group simulated the bundling process using a multi-scale method.[81] In the simulations, upon collision and contact, CNTs "zip" by rapidly aligning parallel to each other, forming a bundle (Fig. 16A and B). This bundle structure is common in CNTs, and occasionally found in agglomerates of SiNWs (Fig. 16C). The simulations show that bundling times increase as a function of collision angle following a power law as a function of CNT length, and that bundling dynamics are dominated by the length of the shorter of the two interacting CNTs.[81] It is proposed that the aerogel forms when the bundling time is of the same order of magnitude as the bundle collision time and in the order of $10^{-1}$ to $10^3$ ns.[81] These results were developed for CNT–CNT collisions but are expected to be applicable to boron nitride nanotubes and NWs.

To compare different FCCVD processes, we previously proposed to analyse the aggregation of 1D nanoparticles in the gas phase through percolation theory, based on the concept of excluded volume.[16] This description relates volumetric concentration ($v$) and aspect ratio ($s$) of the nanoparticles to the probability of forming a continuous network in the volume of interest. For ideal (penetrable) rigid rods, the percolation threshold is $v_p \approx 0.5$ s$^{-1}$. The product $v \cdot s$ then becomes a very rough parameter to assess the likelihood for 1D nanoparticles to form an aerogel in the gas phase, thus marking the boundary between an aerosol and an aerogel. A comparison of this metric for different reports is shown in Fig. 17, including reference values of $v \cdot s$. The graph evidences the wide range of conditions under which FCCVD is operated and their correspondence to the format of the macroscopic material targeted. Low values of $v \cdot s$ are conducive to individualised particles whose aggregation can be controlled downstream to form thin layers for optoelectronics. Mid-range values enable formation of freestanding sheets of nanoparticles randomly oriented in the plane. At the highest end are conditions leading to the formation of a macroscopic aerogel, for example, capable of withstanding continuous drawing from the gas phase as a fibre spinning process.

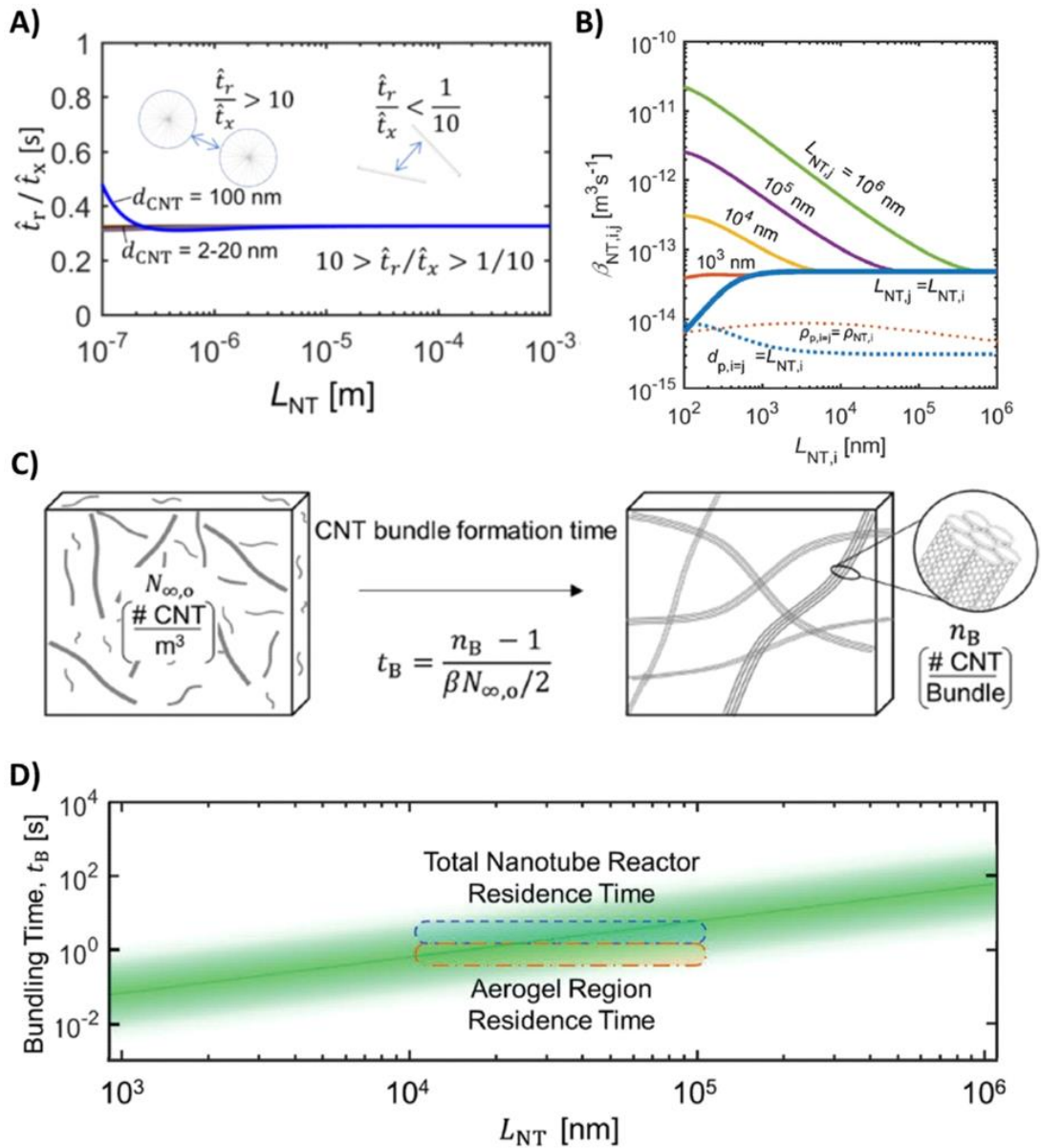

**Fig. 15** Agglomeration and bundling of 1D nanoparticles from collision dynamics. (A) Ratio of rotational to translational characteristic times for different nanoparticle sizes. (B) Dimensional collision kernel for 1D nanoparticles of different length. (C) Schematic and expression to calculate bundling time. (D) Bundling time calculated from the collision kernel compared to experimental residence times in FCCVD growth of CNTs. Reproduced with permission from ref. 79.

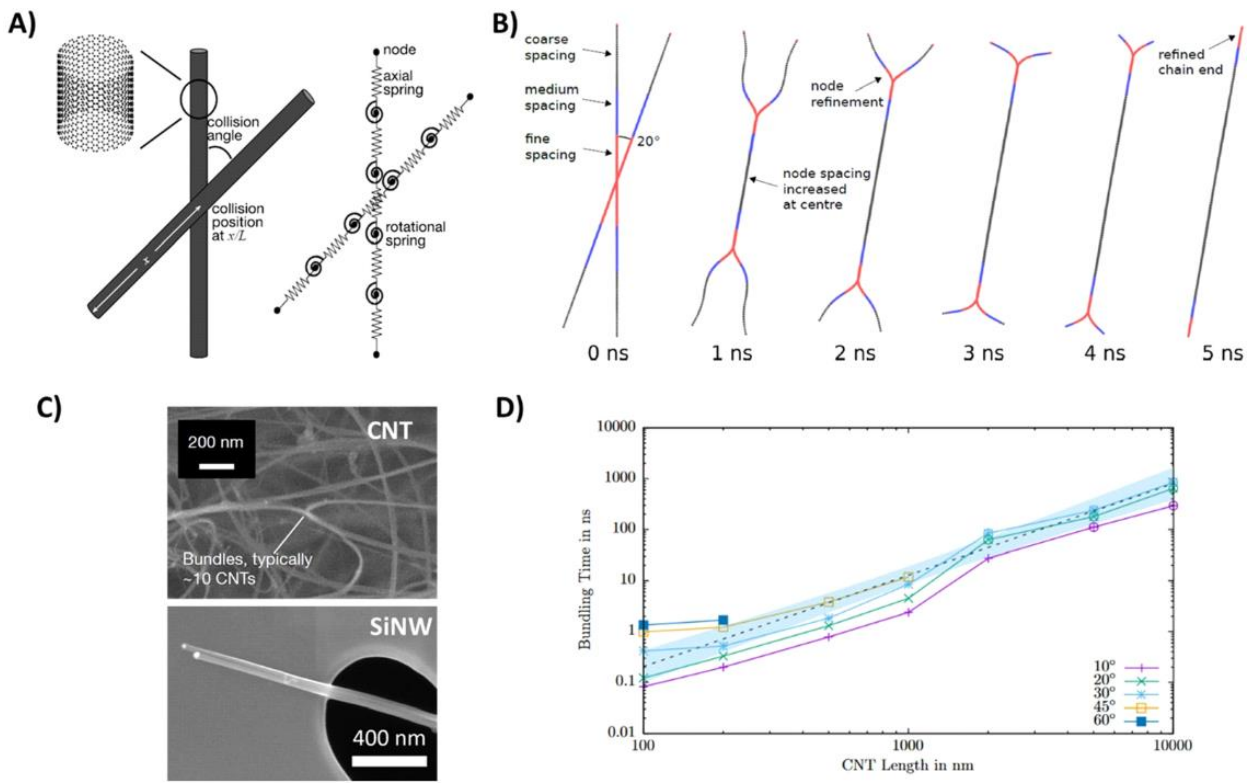

**Fig. 16** Simulation of bundling of CNTs and related 1D nanoparticles. (A) Schematic of two colliding CNTs and the model considering CNTs made of smaller units of cylinders joined together with massless axial and rotational springs. (B) Temporally resolved simulation showing the bundling/ zipping process of two CNTs. (C) Electron micrographs of common bundles of CNTs and occasional bundle-like agglomerates of SiNWs. (D) Bundling time as a function of CNT length obtained from simulations. (A), (B), (C) Top panel and (D) Adapted with permission from 82. Copyright 2020 American Chemical Society. (C) Bottom panel Adapted from ref. 16 with permission from the Royal Society of Chemistry.

# 5. Properties of materials produced by direct assembly: conductors, electrodes, and structural materials

## 5.1 General structure–property relations in nanostructured network materials

In macroscopic ensembles, a longer length of the constituent nanoparticle generally increases bulk properties. For transport properties, reducing the number of ends can reduce the contribution from inter-particle resistance. Longer nanoparticles sustain higher axial stress through a longer length for stress build-up. In fibres of highly aligned CNTs, for example, both longitudinal electrical conductivity and tensile strength have been shown to scale linearly with constituent aspect ratio.[83]

Fig. 18 shows a comparison of literature data on bulk properties of macroscopic ensembles of different CNTs and their corresponding volume fraction. The data show that dispersion-based processing methods, which use CNTs produced by SCVD, are generally constrained to volume fractions below 10 wt%, which limits their bulk properties relative to those of the constituent. Fig. 18 also includes a cluster of data points with high properties, produced from liquid crystal solutions of ≈1 nm-diameter CNTs in strong acids. However, this route is ineffective for larger-diameter nanotubes or nanowires. For reference, the persistence length of a typical inorganic nanowire, a thin CNT and a rigid-rod polymer are >10 000 μm, ≈100 μm[84] and ≈50 nm, respectively.

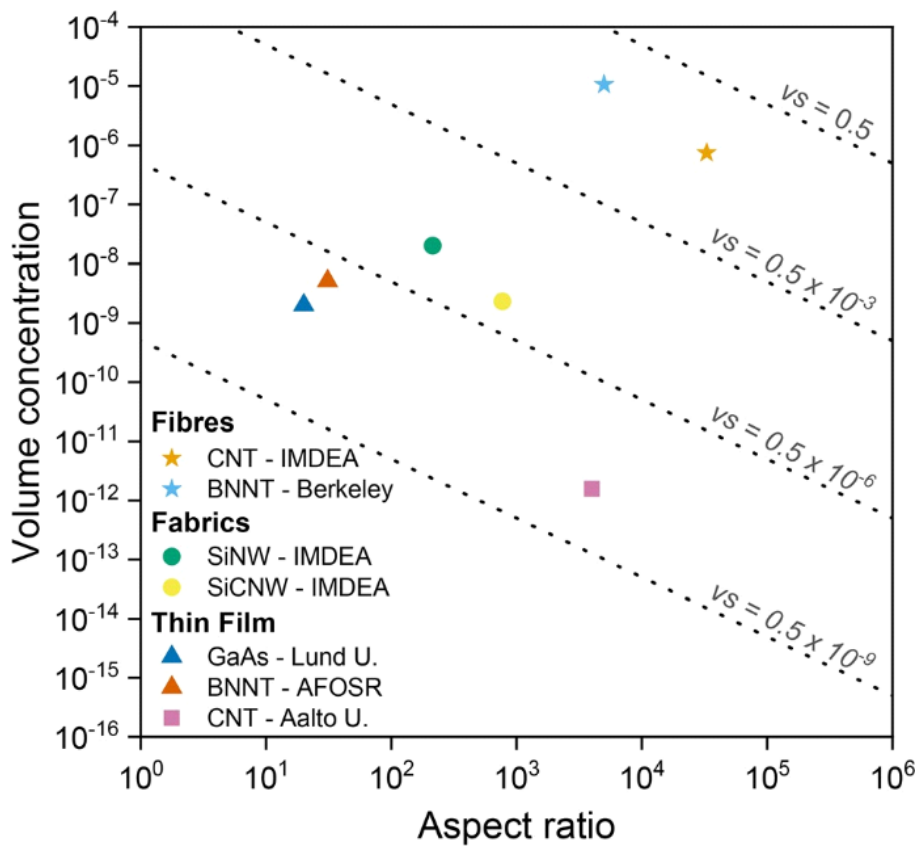

**Fig. 17** Map of different processing regimes used to assemble 1D nanoparticles into macroscopic materials in terms of average nanoparticle aspect ratio ($s$) and nominal volume concentration ($v$).[12–14,16,17,26,56,82] The dashed lines are different products of $v \cdot s$, including the threshold for percolation of ideal (penetrable) rigid rods $v \cdot s = 0.5$.

Assembling high aspect ratio nanoparticles from the gas phase can overcome the limitations of dispersion-based processing. Through gas-phase assembly even CNTs of extraordinarily high aspect ratios above ≈$10^5$ can be assembled into materials with volume fractions of above 20%, leading to bulk properties comparable to the constituents, which compete with best-of-class traditional materials. The medium volume fractions are also relevant for electrodes for energy storage and conversion, as discussed below.

Implicit in the data in Fig. 18 is the importance of alignment in ensembles of high aspect ratio nanoparticles, which largely dominates bulk properties. The low volume fraction systems are comprised of randomly oriented nanoparticles. In such arrangement, they can form freestanding foams, or nanocomposites where the nanoparticles are dispersed in a matrix, often polymeric. Higher volume fraction can be achieved in sheets where the nanoparticles have some degree of alignment through the thickness and random orientation in the plane, analogous to paper sheets or non-woven fabrics. In aligned fibres, 1D nanoparticles are oriented parallel to each other and to the fibre axis, which results in higher packing and volume fractions approaching 100%. Fig. 19 shows examples of different materials with the three types of orientation, all produced through gas-phase assembly. The SiNW sheet and aligned CNT fibre have ≈25% and ≈65% of the density of their constituent crystal, respectively. For reference, in those formats, they have values of $v \cdot s > 50$, showing the significant densification from the aerogel state.

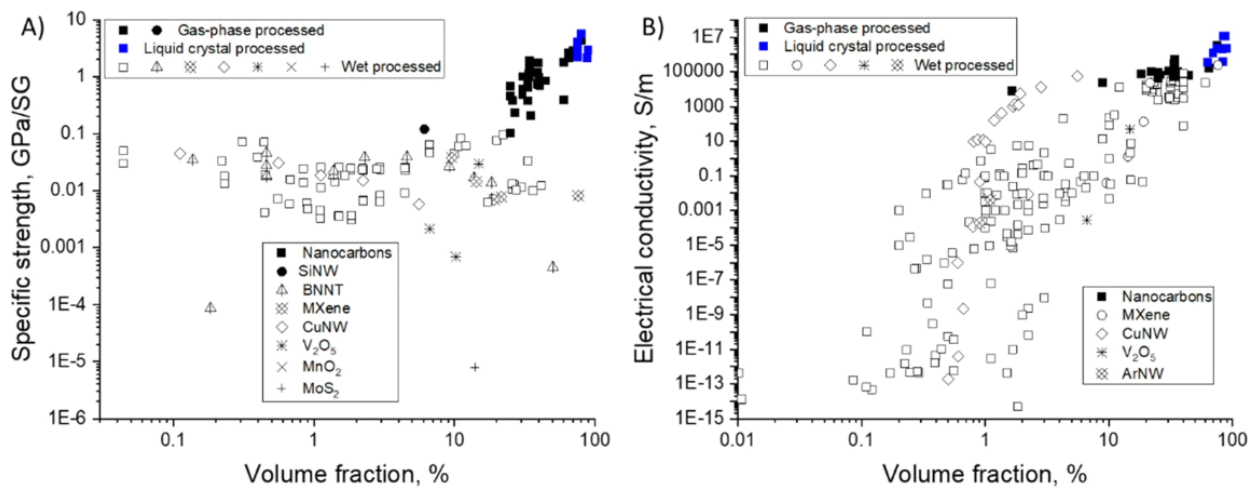

**Fig. 18** Bulk properties (A) Specific strength and (B) Electrical conductivity and volume fraction of high-aspect ratio nanoparticle ensembles produced by different methods. Volume fraction is considered with respect to the whole volume of the sample, including the pores.

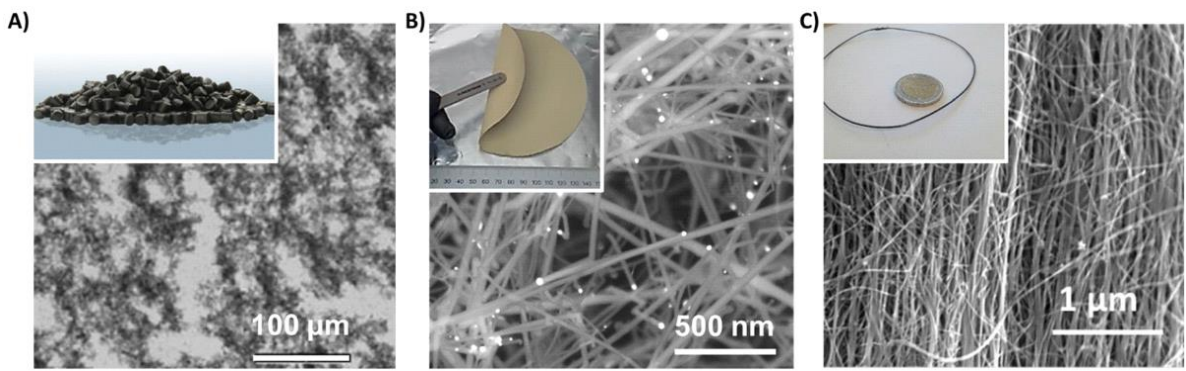

**Fig. 19** Network structure of nanowire ensembles. (A) CNT dispersion in the polymer matrix (B) Sheet of SiNWs predominantly aligned through the thickness. (C) Fibre of CNTs aligned parallel to the fibre axis. (A) Adapted with permission from ref. 85. Copyright 2011 Springer Nature and Copyright 2022 Arkema. (C) Adapted with permission from ref. 63 and 132. Copyright 2021 and 2019, respectively, Elsevier.

## 5.2 Nanowire network materials with textile-like properties

The macroscopic ensembles of 1D nanoparticles assembled from the gas phase have inherently a network structure, consisting of weakly interacting, aggregated long nanoparticles resembling a nanotextile. Indeed, they have textile-like properties, such as flexibility in bending, resistance to cutting and the possibility for weaving, knitting,[86] knotting,[87] *etc*. Overall, this network structure implies that their mechanical behaviour is often radically different from that of their constituents in the monolithic state. The most evident example is their ability to sustain large deformations. Under tensile deformation, for example, sheets of different NWs reach strain-to-break in excess of 2.5%, compared to the 0.1–1% for their monolithic analogues, such as boron fibres (BF), carbon fibres (CF) and bulk silicon (Fig. 20A). These sheets have a characteristic sailshaped stress–strain curve resulting from combined elastoplastic deformation and progressive sliding of elements. This behaviour contrasts with the brittle, linear-elastic deformation of monolithic analogues and is instead similar to that of paper and other athermal fibrillary networks.[88]

The deformation mechanism in nanostructured networks can lead to much larger damage-tolerance than in monolithic analogues, as crack propagation is arrested and diverted laterally. Given the relatively large and elastoplastic nature of tensile deformation of these networks, their damage tolerance can be best characterised in terms of work of fracture, *i.e.*, the energy required to break them, instead of fracture toughness. Aligned fibres of CNTs produced from the gas phase have work of fracture and strength, a combination of interest for impact, which is above traditional high-performance synthetic fibres (Fig. 20B). Gas phase-produced unidirectional fabrics of CNTs have work of fracture also above that of CF, a property that contributes to preserving their properties upon recycling from structural composites. Nanostructured networks of SiNWs have a work of fracture of 0.36 J g$^{-1}$, compared with the 0.014 J g$^{-1}$ of their monolithic analogues and orders of magnitude above that of particulate electrodes (0.003 J g$^{-1}$),[16] for example. The emerging evidence is that gas-phase processing into nanostructured networks imbues inorganic materials with a "toughness" otherwise inaccessible in the particulate or monolithic format.

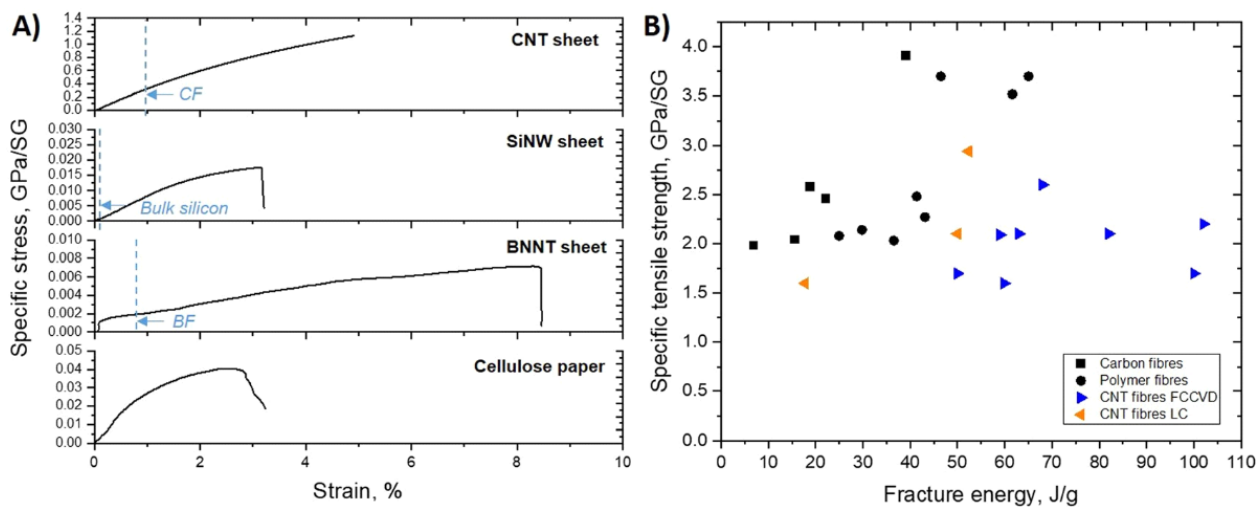

**Fig. 20** Tensile properties of sheets of 1D nanoparticles compared to conventional materials. (A) Stress–strain curves for sheets of different NWs (BNNTs,[29] SiNWs[16] and CNTs[89]) compared to office paper[90] and to their constituents in monolithic form: boron fibre (BF),[91] bulk Si and CF. (B) Map of specific tensile strength and fracture energy showing that fibres of aligned CNTs have combined properties beyond the best synthetic fibres.

## 5.3 Electrical conductors

Gas-phase assembly has been particularly successful in the manufacture of transparent conductors. The general process consists in the direct deposition of a thin network of CNTs onto a transparent substrate (Fig. 21). Industrial-scale transparent conductors based on (doped) SWCNTs with 90% transparency in the visible range have reached a sheet resistance of 35 Ohm sq$^{-1}$,[24] significantly below the 100 Ohm sq$^{-1}$ resistance of the conventional ITO standard. The high conductivity of these networks stems from the long length of the constituent CNTs, control over their diameter and aggregation into bundles,[24] and the introduction of dopants.[92] Here again, gas-phase synthesis stands out over wet-processing by enabling processing of long nanoparticles with controlled aggregation as engineered networks. The figure of merit for transparent conductors is the ratio of optical $\sigma_{op}$ to bulk DC conductivity $\sigma_{dc,B}$, which is related to transmittance ($T$) and sheet resistance ($R_s$) by $T = \left(1 + \frac{Z_0}{2R_s}\frac{\sigma_{op}}{\sigma_{dc,B}}\right)^{-2}$, where $Z_0$ is the impedance of free space. As shown in Fig. 21E, the figure of merit for CNTs is significantly higher for gas-phase assembly than for wet processing.

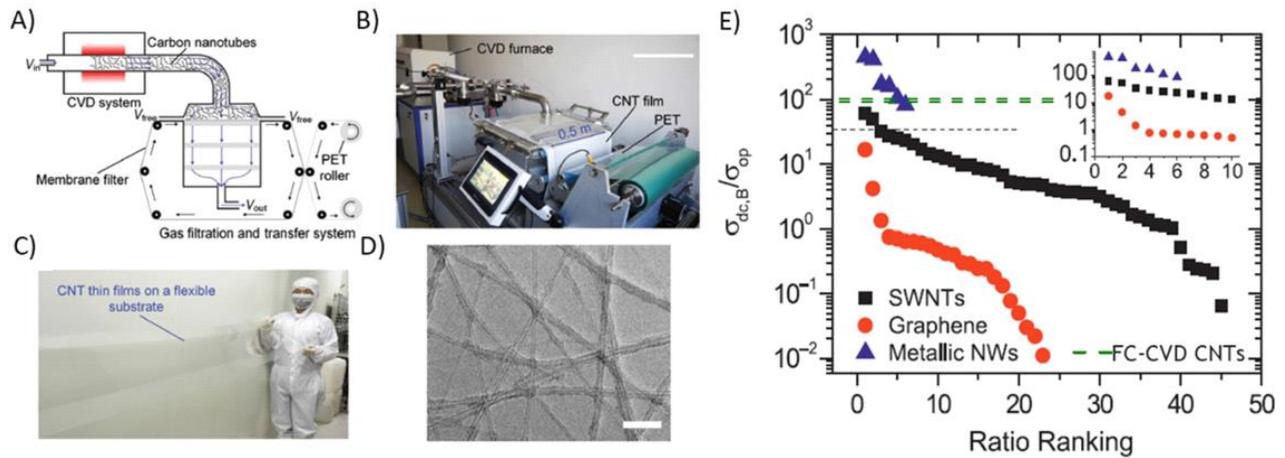

**Fig. 21** Transparent conductors of CNTs. (A) Scheme and photograph (B) of the FCCVD manufacturing process. (C) Example of a scale-up transparent conductor and (D) TEM micrograph showing the SWCNT network structure. (E) Figure of merit for transparent conductors of CNTs produced by wet-processing[95] and gas-phase assembly[11,93] (green dashed lines), including other materials for reference. (A-D) Adapted with permission from ref. 94. Copyright 2021 Wiley-VCH. (E) Modified with permission from ref. 95. Copyright 2011 Springer Nature.

## 5.4 Nanostructured electrodes

The use of gas phase assembly to produce electrodes is part of a broader trend to explore architectures beyond granular electrodes with a layered structure supported on a metallic foil (Fig. 22). Replacing granular materials (active and conducting additives) with a nanostructured network of high aspect ratio materials can lead to composite architectures that improve the performance of electrodes, open new electrode processing routes, and enable a myriad of complex-shaped and deformable devices.[96,97]

An instructive example is battery electrodes where particulate conductors are replaced by high aspect ratio CNTs. At mass fractions above ≈1 wt%, SWNTs can form a continuous (segregated) network that contains the active material and forms a composite structure even without polymer binders.[98] Out-of-plane electrical conductivity and tensile strength are two important electrode properties dominated by the CNT network, which led to a comparison of different processing methods (Fig. 23).

Pioneering work on composite electrodes produced from slurries of SWNTs and granular active materials shows that their out-of-plane conductivity follows a percolation model, with a limiting value of ≈0.1 S m$^{-1}$,[99] which is close to the proposed threshold (1 S m$^{-1}$) to prevent electron transport limitations in batteries (Fig. 23A). Using different active materials, these composites showed strength above 4 MPa, enabling manufacture of thick electrodes that overcome the limiting critical crack thickness of particulate coatings of around 100 microns.[98]

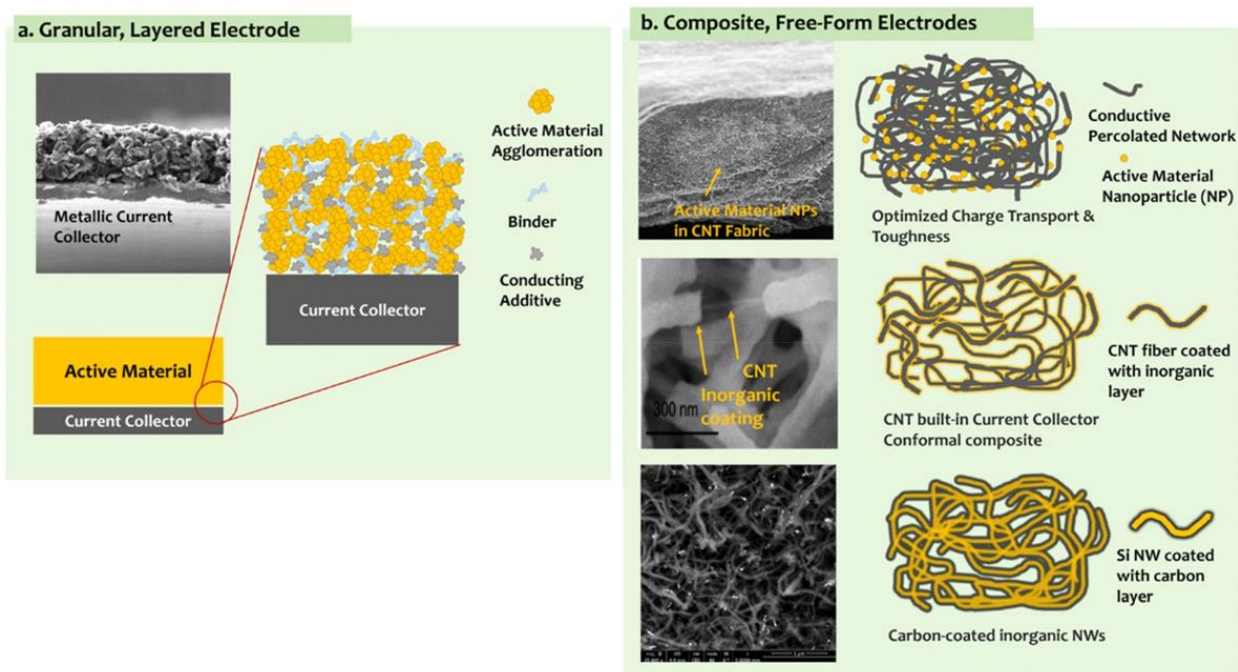

**Fig. 22**  Schematic comparison of conventional and composite electrodes produced through gas phase processing. (a) Granular electrodes of active material on a metallic foil have a layered structured. (b) Electrodes based on networks of high aspect ratio nanoparticles resemble a composite and can eliminate the need for metallic current collectors. Micrograph adapted with permission from ref. 110. Copyright 2021 American Chemical Society.

In an example of a gas phase method, CNTs are directly grown on spherical active particles flowing through a CVD reactor. The "hairy" particles are then processed as the conventional granular active material to form electrodes deposited on a current collector. Electrodes produced from hairy particles reach 10–50 wt% of CNTs and higher conductivity and superior electrochemical performance compared to control samples.[100]

CNT nanotextiles produced by FCCVD have been widely used to produce a variety of such composite electrodes, whereby the active material is incorporated into the pre-made CNT nanotextile scaffold.[101] $MoS_2$/CNT nanotextiles have shown out-of-plane electrical conductivity above the threshold for electronic rate-limitations in charge/discharge of LIBs. They also have specific tensile strength higher than that of steel and over an order of magnitude higher than that of wet-processed analogues (Fig. 23B). The electrical and mechanical properties of these composite electrodes stem from the long length of CNTs, but also from a higher mass fraction. A comparison of the composites from wet- and gas-phase processing shows that they are generally in different mass fraction ranges. The limited data available also suggest that bulk properties in these systems scale differently with mass fraction (Fig. 23B). There are many other examples of improved performance in electrodes with gas phase-assembled processed CNT networks, including stabilisation of battery active materials,[102–104] and increased conversion efficiency in various photo(electro)catalytic electrodes.[105–109] These processes are, however, more strongly dependent on the active material and multiple interfaces in the system and thus harder to compare against wet-processed analogues.

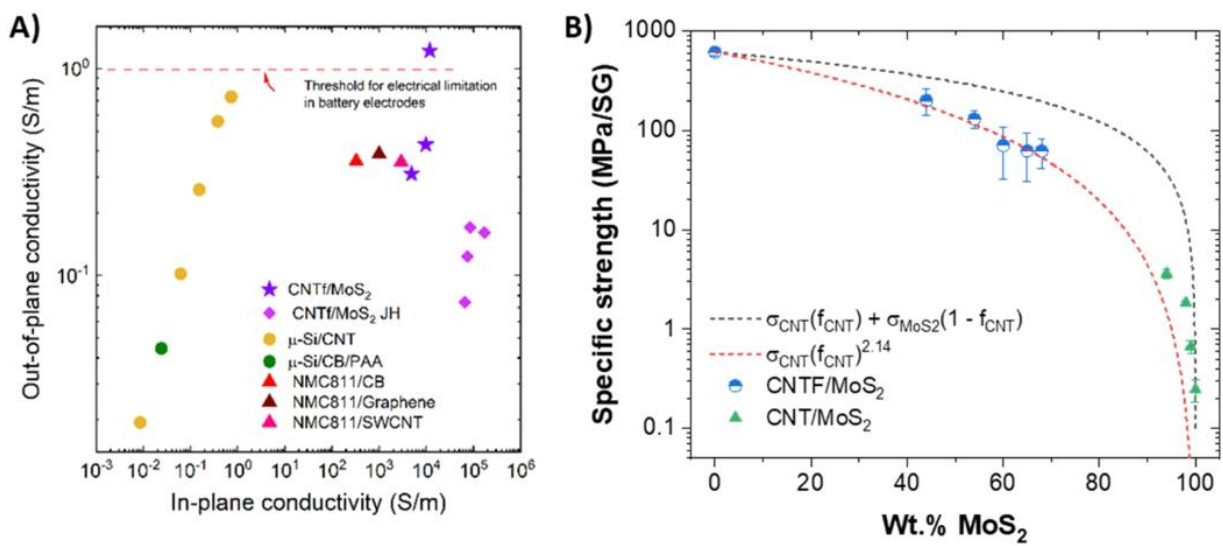

**Fig. 23** Electrical and mechanical properties of composite electrodes based on FCCVD-synthesised CNT nanotextiles and wet-processed nanocarbons, respectively. (A) Electrode out-of-plane and in-plane electrical conductivity.[98,99,110] (B) Specific tensile strength for different mass fractions of active material.[111,112] Adapted with permission from ref. 113. Copyright 2023 American Chemical Society.

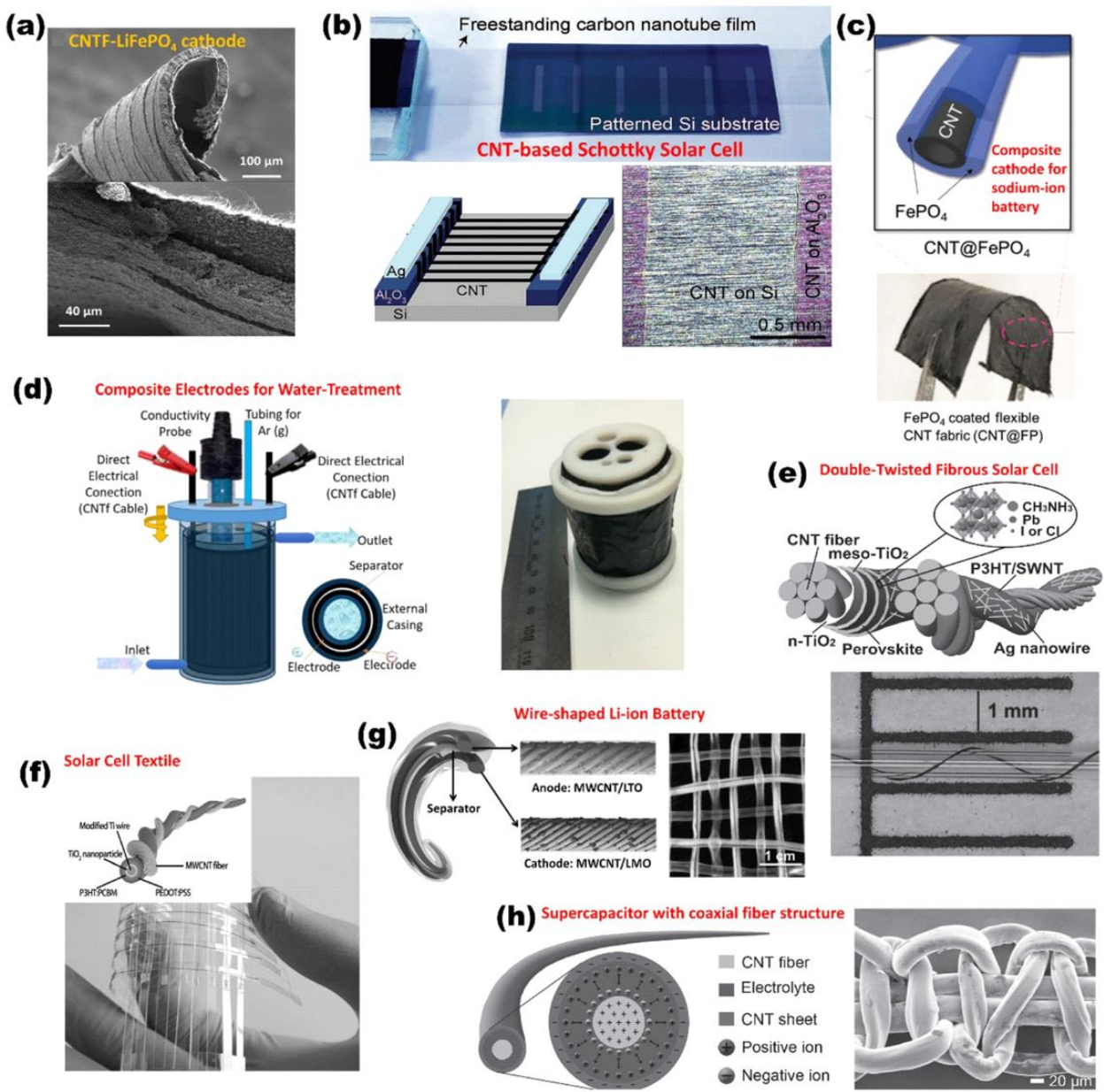

**Fig. 24** Complex-shaped and deformable devices based on composite electrodes with a backbone of CNT nanotextiles produced by FCCVD.[113,115,120–125] (A) Adapted with permission from ref. 113. Copyright 2019 American Chemical Society. (B) Adapted with permission from ref. 120. Copyright 2013 Wiley-VCH. (C) Adapted with permission from ref. 121. Copyright 2018 Wiley-VCH. (D) Adapted with permission from ref. 115. Copyright 2021 Elsevier. (E) Adapted with permission from ref. 122. Copyright 2015 Wiley-VCH. (F) Adapted with permission from ref. 123. Copyright 2014 Wiley-VCH. (G) Adapted with permission from ref. 124. Copyright 2014 Wiley-VCH. (H) Adapted with permission from ref. 125. Copyright 2013 Wiley-VCH.

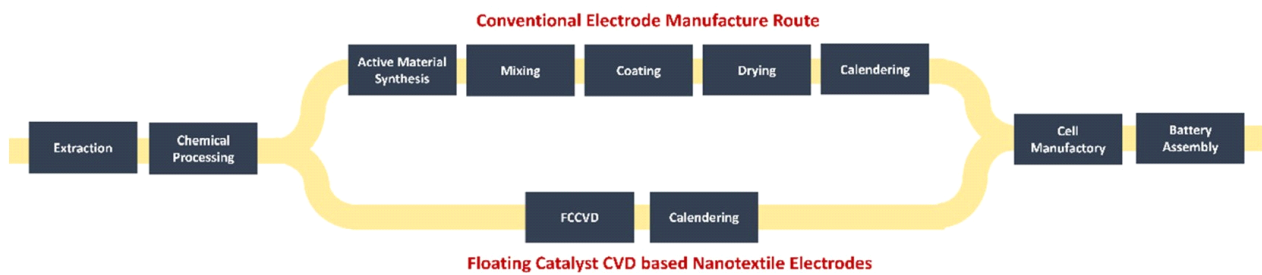

**Fig. 25** Schematic of the stages in traditional and FCCVD nanotextile fabrication of electrodes and their assembly in battery cells.[128]

Given their combined strength and longitudinal conductivity at high mass fractions, the nanocarbon networks in composite electrodes can act as a built-in current collector, eliminating metallic foil current collectors and enabling electrodes with complex shapes. Examples of non-conventional electrode formats include fibre-shaped,[96] sheet-like,[113] and transparent,[114] for multiple energy storage,[115] conversion[116] and energy harvesting processes.[117] Some examples are included in Fig. 24. The underlying feature is the relatively high tensile strength and strain-to-break of the nanocarbon network backbone, combined with a small device thickness. In conjunction with a suitable soft matrix, patterning of CNT nanotextile current collectors has been used to demonstrate multiple types of stretchable opto(electronic) devices.[118] With a stiff matrix and special electrolytes, they have formed structural energy-storing composites.[119] These examples illustrate the plethora of architectures and devices enabled by replacement of metallic foil current collectors with FCCVD-produced CNT nanotextiles.

An emerging area of interest is the synthesis of high aspect ratio active materials directly in the gas phase. After all, many of today's particulate battery materials are produced industrially through thermal synthesis routes in flow-through reactors, including established active materials and conductors, and a range of emerging nanoparticles for anodes and cathodes.[126,127] These particles are then dispersed in suitable solvents, mixed with a binder to form a slurry and coated on a current collector to form electrodes. Recently, we demonstrated that high aspect ratio SiNWs produced by FCCVD can be directly assembled as electrodes from the gas-phase, eliminating all need for solvents and mixing steps.[36] In its nanotextile form, the active material is freestanding, flexible and has sufficient mechanical robustness for downstream processing into battery cells, altogether avoiding slurry processing. The implications are beyond Si anodes and LIBs. FCCVD growth of active materials opens a completely new route for the manufacture of electrodes through gas-to-nanotextile processing (Fig. 25).

Extending the application of FCCVD beyond CNT conductors and SiNW anodes is an appealing prospect. The typical volume fraction of porous materials in electrodes for energy storage (*e.g.*, active material[129]), photocatalysis (*e.g.*, $TiO_2$ photoelectrode[130]) and electrocatalysis (catalyst supports and gas diffusion layers[131]) is 25–80%. These volume fractions are high to attain through wet processing, but are readily accessible through gas phase assembly. From a more fundamental point of view, the electrodes comprised exclusively of an active material as a nanostructured network are interesting model systems that can help understand the interfacial role of binders and conductive additives, the contributions of inter- and intra-particle charge transport, and the contribution of different rate-limiting electrochemical processes.

From the point of view of applications, as 1D nanoparticles become increasingly available as macroscopic nanotextiles, there will be more focus on further processing them into components. The associated challenges to interface nanotextiles with polymers to make composites[132,133] or with metals[134–136] for charge transfer process, for example, have been largely addressed for CNT nanotextiles and led to methods transferable to nanotextiles of other chemical compositions.

## 6. Future directions

As discussed throughout this review, transforming gas/liquid precursors into densified nanotextiles directly as they are synthesised in the gas phase enables processing of high aspect ratio nanoparticles into macroscopic materials with high-performance properties. However, as multiple transformation steps are integrated in a single manufacturing process, it becomes necessary to introduce on-line measurement methods to characterise individual processing steps. This will require different experimental methods to monitor the composition and morphology during 1D nanoparticle growth, aggregation and densification. Similarly, the process as a whole would benefit from development of models that bridge different length scales of materials and time scales of processes involved in the three main stages of the manufacturing route.

Concerning primarily the growth stage, there is a need for fundamental studies that describe the reaction mechanisms and kinetics of 1D nanoparticle growth in the FCCVD regime and which can provide a complete description of its inherently fast growth rate. A better understanding on the fundamental factors that govern selectivity will accelerate the synthesis and assembly of other 1D nanoparticles, a pressing need considering that there are roughly 10 times more 1D nanoparticles producible by FCCVD (>70) than the total ever synthesised by this method (5) over the last 40 years. Improvements are also necessary to understand the reaction mechanism and underlying kinetics in order to limit competing reaction paths that deplete the reagents and contaminate the 1D nanoparticle product, particularly the formation of soot, in order to preserve selectivity while increasing reaction throughput.

In addition to increasing the palette of ultra-long 1D nanoparticles produced by FCCVD, future efforts should be directed at controlling their crystal structure and purity. This is a significant challenge given the extremely short reaction times involved, the mobility and evolution of the catalyst particles in the gas-phase, and the inherent breadth of catalyst size distribution. A promising avenue is to perform *in situ* spectroscopic measurements during growth of 1D nanoparticles, particularly under resonance conditions that therefore allow detection at low concentrations.

The aggregation of high aspect ratio nanoparticles in the gas phase is central to this manufacturing route, yet it is the least understood stage. Particularly because of the lack of analogous processes in the scientific realm and the inherent technical challenges in the detection of aerosols and aerogels in the FCCVD environment. These are invitations to explore a fertile ground for development of further instrumentation and modelling. An improved understanding of aerogel formation is likely to benefit from studying different 1D nanoparticles that span a range of morphologies and aggregation dynamics. It will require the implementation and development of multiple techniques to characterise particles and ensembles on different length-scales, both *in situ* and offline. This involves not only integrating characterisation tools in the 1D nanoparticle reactors, but also developing portable model reactors to high intensity synchrotron facilities.

Advances are also sought in describing three-dimensional bulk mechanical and transport properties of nanotextiles, including those with a moderate degree of alignment. Building on existing micro-mechanical models, a future goal should be to describe the large strain-to-break and intrinsic damage-tolerance of these network materials. These properties set them apart from their monolithic analogues and can have important implications for extended durability and re-use of structural materials and electrodes.

# Author contributions

I. G. P.: conceptualisation, formal analysis, investigation, data curation, writing, editing, visualisation; M. V. P.: conceptualisation, formal analysis, investigation, data curation, writing, editing, visualisation; R. S. S.: conceptualisation, formal analysis, investigation, data curation; A. M.: conceptualisation, formal analysis, investigation, data curation, writing, editing, visualisation; A. P.: conceptualisation, formal analysis, investigation, data curation, writing, editing, visualisation; A. R.: formal analysis, editing, funding acquisition; J. J. V.: conceptualisation, formal analysis, investigation, writing, editing, project administration, funding acquisition.

# Conflicts of interest

R. S. and J. J. V. have a financial interest in Floatech, a company that commercialises SiNW anodes produced by FCCVD.

# Appendix

Glossary

**1D nanoparticle**: nanoparticle[137] with an aspect ratio above ≈100 with a nanotube or nanowire structure.

**Aerosol**: suspension of nanoparticles in the gas phase. The aerosol may also comprise aggregates of a few 1D nanoparticles.

**Percolation:** formation of a continuous network of aggregated 1D nanoparticles in a macroscopic volume.

**1D nanoparticle aerogel**: percolated network of 1D nanoparticles, with ultra-low mass fraction or bulk density ($\rho_{bulk}/\rho_{1D\ NP}$ < 0.01), mechanically stable against the stresses generated from the flow and collection of materials in FCCVD.

**Nanotextile**: a class of materials consisting of a network of high aspect ratio nanoparticles, characterised by a significant porosity (>≈10%), and interaction between nanoparticles through secondary bonds.

**Sheet:** example of freestanding nanotextiles with random orientation in the plane, sometimes also referred to as mats. For CNTs, this was traditionally referred to as bucky paper. (In case where the 1D nanoparticles have some degree of alignment in the plane, for example when made of wound CNT fibre aerogels, and because of the large size of components routinely produced in industry, *e.g.*, 1 m wide and >10 m long, these are more appropriately referred to as non-woven fabrics.) Soot: nanoscale material formed through the direct pyrolysis of precursors without intervention of the catalyst during growth, typically in the form of quasi-spherical nanoparticles or clusters that may attach to nanowires.

# Acknowledgements

This work was supported by the European Union Horizon 2020 Programme under grant agreement 101045394 (ERC-2021-COG, UNIYARNS), Marie Sklodowska Curie Fellowship SUPERYARN under grant number 101029091, by the Madrid Regional Government (FOTOART-CMP2018/ NMT-4367 and UPM/APOYO-JOVENES-F6TCCN-145-GGT34M), (MAD2D-CM)-MRR MATERIALES AVANZADOS-IMDEA, and by The Carbon Hub.

# References

1. P. Potschke, T. D. Fornes and D. R. Paul, *Polymer*, 2002, 43, 3247–3255.
2. Hardy, J. Dix, C. D. Williams, F. R. Siperstein, P. Carbone and H. Bock, *ACS Nano*, 2018, 12, 1043–1049.
3. Backes, T. M. Higgins, A. Kelly, C. Boland, A. Harvey, D. Hanlon and J. N. Coleman, *Chem. Mater.*, 2017, 29, 243–255.
4. M. Endo, M. Shikata, M. Momose and T. Shiraishi, 17th Biennial Carbon Conference Proceedings, 1985, pp. 295–296.
5. M. Endo, *Chemtech*, 1988, 18, 568–576.
6. Moisala, A. G. Nasibulin and E. I. Kauppinen, *J. Phys.: Condens. Matter*, 2003, 15, S3011–S3035.
7. M. J. Bronikowski, P. A. Willis, D. T. Colbert, K. A. Smith and R. E. Smalley, *J. Vac. Sci. Technol., A*, 2001, 19, 1800–1805.
8. H. W. Zhu, C. L. Xu, D. H. Wu, B. Q. Wei, R. Vajtai and P. M. Ajayan, *Science*, 2002, 296, 884–886.
9. Y.-L. Li, I. A. Kinloch and A. H. Windle, *Science*, 2004, 304, 276 LP–278 LP.
10. X.-H. Zhong, Y.-L. Li, Y.-K. Liu, X.-H. Qiao, Y. Feng, J. Liang, J. Jin, L. Zhu, F. Hou and J.-Y. Li, *Adv. Mater.*, 2010, 22, 692–696.
11. Q. Zhang, W. Zhou, X. Xia, K. Li, N. Zhang, Y. Wang, Z. Xiao, Q. Fan, E. I. Kauppinen and S. Xie, *Adv. Mater.*, 2020, 32, 1–8.
12. J. K. Myung, S. Chatterjee, M. K. Seung, E. A. Stach, M. G. Bradley, M. J. Pender, L. G. Sneddon and B. Maruyama, *Nano Lett.*, 2008, 8, 3298–3302.
13. M. Heurlin, M. H. Magnusson, D. Lindgren, M. Ek, L. R. Wallenberg, K. Deppert and L. Samuelson, *Nature*, 2012, 492, 90–94.
14. Fathalizadeh, T. Pham, W. Mickelson and A. Zettl, *Nano Lett.*, 2014, 14, 4881–4886.
15. W. Metaferia, A. R. Persson, K. Mergenthaler, F. Yang, W. Zhang, A. Yartsev, R. Wallenberg, M. E. Pistol, K. Deppert, L. Samuelson and M. H. Magnusson, *Nano Lett.*, 2016, 16, 5701–5707.
16. R. S. Schäufele, M. Vazquez-Pufleau and J. J. Vilatela, *Mater. Horiz.*, 2020, 7, 2978–2984.
17. I. Gómez-Palos, M. Vazquez-Pufleau, J. Valilla, Á. Ridruejo, D. Tourret and J. J. Vilatela, *Nanoscale*, 2022, 14, 18175–18183.
18. K. Koziol, J. Vilatela, A. Moisala, M. Motta, P. Cunniff, M. Sennett and A. Windle, *Science*, 2007, 318, 1892–1895.
19. Y. S. Cho, J. W. Lee, J. Kim, Y. Jung, S. J. Yang and C. R. Park, *Adv. Sci.*, 2022, 2204250.
20. L. Weller, F. R. Smail, J. A. Elliott, A. H. Windle, M. Boies and S. Hochgreb, *Carbon*, 2019, 146, 789–812.
21. B. Mas, B. Alemán, I. Dopico, I. Martin-Bragado, T. Naranjo, E. M. Pérez and J. J. Vilatela, *Carbon*, 2016, 101, 458–464.
22. S. H. Lee, J. Park, J. H. Park, D. M. Lee, A. Lee, S. Y. Moon, S. Y. Lee, H. S. Jeong and S. M. Kim, *Carbon*, 2021, 173, 901–909.
23. Y. Liao, H. Dong, Q. Zhang, N. Wei, E.-X. Ding, S. Ahmad, H. Jiang and E. I. Kauppinen, *Diamond Relat. Mater.*, 2021, 120, 108716.
24. Q. Zhang, J.-S. Nam, J. Han, S. Datta, N. Wei, E.-X. Ding, Hussain, S. Ahmad, V. Skakalova, A. T. Khan, Y.-P. Liao, M. Tavakkoli, B. Peng, K. Mustonen, D. Kim, I. Chung, S. Maruyama, H. Jiang, I. Jeon and E. I. Kauppinen, *Adv. Funct. Mater.*, 2022, 32, 2103397.
25. D. Sun, M. Y. Timmermans, Y. Tian, A. G. Nasibulin, E. I. Kauppinen, S. Kishimoto, T. Mizutani and Y. Ohno, *Nat. Nanotechnol.*, 2011, 6, 156–161.
26. S. Chatterjee, M. J. Kim, D. N. Zakharov, S. M. Kim, E. A. Stach, B. Maruyama and L. G. Sneddon, *Chem. Mater.*, 2012, 24, 2872–2879.
27. J. H. Kim, T. V. Pham, J. H. Hwang, C. S. Kim and M. J. Kim, *Nano Converg.*, 2018, 5, 17.
28. C. Zhi, Y. Bando, C. Tan and D. Golberg, *Solid State Commun.*, 2005, 135, 67–70.

29. P. Nautiyal, C. Zhang, A. Loganathan, B. Boesl and Agarwal, *ACS Appl. Nano Mater.*, 2019, 2, 4402–4416.
30. M. B. Jakubinek, B. Ashrafi, Y. Martinez-Rubi, J. Guan, M. Rahmat, K. S. Kim, S. Dénommée, C. T. Kingston and B. Simard, in Nanotube Superfiber Materials, ed. M. J. Schulz, V. Shanov, Z. Yin and M. B. T.-N. S. M. Second E. Cahay, Elsevier, 2019, pp. 91–111.
31. W. Zhang, F. Yang, M. E. Messing, K. Mergenthaler, M. E. Pistol, K. Deppert, L. Samuelson, M. H. Magnusson and A. Yartsev, *Nanotechnology*, 2016, 27, 1–6.
32. S. Sivakumar, A. R. Persson, W. Metaferia, M. Heurlin, R. Wallenberg, L. Samuelson, K. Deppert, J. Johansson and M. H. Magnusson, *Nanotechnology*, 2021, 32, 025605.
33. M. T. Borgström, M. H. Magnusson, F. Dimroth, G. Siefer, O. Höhn, H. Riel, H. Schmid, S. Wirths, M. Björk, I. Åberg, W. Peijnenburg, M. Vijver, M. Tchernycheva, V. Piazza and L. Samuelson, *IEEE J. Photovolt.*, 2018, 8, 733–740.
34. E. Barrigón, O. Hultin, D. Lindgren, F. Yadegari, M. H. Magnusson, L. Samuelson, L. I. M. Johansson and M. T. Björk, *Nano Lett.*, 2018, 18, 1088–1092.
35. M. H. Magnusson, B. J. Ohlsson, M. T. Björk, K. A. Dick, M. T. Borgström, K. Deppert and L. Samuelson, *Front. Phys.*, 2014, 9, 398–418.
36. M. Rana, A. Pendashteh, R. S. Schäufele, J. Gispert and J. J. Vilatela, *Adv. Energy Mater.*, 2022, 12, 2103469.
37. V. G. Dubrovskii and Y. Y. Hervieu, *J. Cryst. Growth*, 2014, 401, 431–440.
38. V. G. Dubrovskii, N. V. Sibirev, J. C. Harmand and F. Glas, *Phys. Rev. B*, 2008, 78, 235301.
39. J. Johansson and M. H. Magnusson, *J. Cryst. Growth*, 2019, 525, 125192.
40. G. G. Tibbetts, C. A. Bernardo, D. W. Gorkiewicz and R. L. Alig, *Carbon*, 1994, 32, 569–576.
41. R. Rao, N. Pierce, D. Liptak, D. Hooper, G. Sargent, S. L. Semiatin, S. Curtarolo, A. R. Harutyunyan and B. Maruyama, *ACS Nano*, 2013, 7, 1100–1107.
42. J. Kikkawa, Y. Ohno and S. Takeda, *Appl. Phys. Lett.*, 2005, 86, 1–3.
43. V. Dhaka, T. Haggren, H. Jussila, H. Jiang, E. Kauppinen, T. Huhtio, M. Sopanen and H. Lipsanen, *Nano Lett.*, 2012, 12, 1912–1918.
44. M. R. Ramdani, E. Gil, C. Leroux, Y. André, A. Trassoudaine, D. Castelluci, L. Bideux, G. Monier, C. Robert-Goumet and R. Kupka, *Nano Lett.*, 2010, 10, 1836–1841.
45. C. Tang, Y. Bando, T. Sato and K. Kurashima, *Chem. Commun.*, 2002, 2, 1290–1291.
46. C. Tang, Y. Bando and T. Sato, *Chem. Phys. Lett.*, 2002, 362, 185–189.
47. S. E, L. Wu, C. Li, Z. Zhu, X. Long, R. Geng, J. Zhang, Z. Li, W. Lu and Y. Yao, *Nanoscale*, 2018, 10, 13895–13901.
48. Pakdel, C. Zhi, Y. Bando, T. Nakayama and D. Golberg, *Nanotechnology*, 2012, 23, 215601.
49. Y. Huang, J. Lin, C. Tang and Y. Bando, *Nanotechnology*, 2011, 22, 145602.
50. C. H. Lee, M. Xie, V. Kayastha, J. Wang and Y. K. Yap, *Chem. Mater.*, 2010, 22(5), 1782–1787.
51. B. Park, Y. Ryu and K. Yong, *Surf. Rev. Lett.*, 2004, 11, 373– 378.
52. J. A. Rajesh and A. Pandurangan, *J. Nanosci. Nanotechnol.*, 2014, 14, 2741–2751.
53. S. K. Panda, J. Sengupta and C. Jacob, *J. Nanosci. Nanotechnol.*, 2010, 10, 3046–3052.
54. G. Attolini, F. Rossi, M. Negri, S. C. Dhanabalan, M. Bosi, F. Boschi, P. Lagonegro, P. Lupo and G. Salviati, *Mater. Lett.*, 2014, 124, 169–172.
55. H. Y. Kim, J. Park and H. Yang, *Chem. Commun.*, 2003, 256–257.
56. V. Reguero, B. Alemán, B. Mas and J. J. Vilatela, *Chem. Mater.*, 2014, 26, 3550–3557.
57. M. Vazquez-Pufleau and M. Yamane, *Chem. Eng. Sci.*, 2020, 211, 115230.
58. R. S. Schäufele, M. Vazquez-Pufleau, A. Pendashteh and J. J. Vilatela, *Nanoscale*, 2022, 14, 55–64.
59. X. Rodiles, V. Reguero, M. Vila, B. Alemán, L. Arévalo, F. Fresno, V. A. de la Peña O'Shea and J. J. Vilatela, *Sci. Rep.*, 2019, 9, 9239.
60. K. S. Kim, C. T. Kingston, A. Hrdina, M. B. Jakubinek, J. Guan, M. Plunkett and B. Simard, *ACS Nano*, 2014, 8, 6211–6220.
61. B. Alemán, R. Ranchal, V. Reguero, B. Mas and J. J. Vilatela, *J. Mater. Chem. C*, 2017, 5, 5544–5550.

62. M. S. Motta, A. Moisala, I. A. Kinloch and A. H. Windle, *J. Nanosci. Nanotechnol.*, 2008, 8, 2442–2449.
63. Mikhalchan, M. Vila, L. Arévalo and J. J. Vilatela, *Carbon*, 2021, 179, 417–424.
64. Alemán, M. M. Bernal, B. Mas, E. M. Pérez, V. Reguero, G. Xu, Y. Cui and J. J. Vilatela, *Nanoscale*, 2016, 8, 4236–4244.
65. M. Vazquez-Pufleau, I. Gomez-Palos, L. Arévalo, J. GarcíaLabanda and J. J. Vilatela, *Adv. Powder Technol.*, 2023, 34, 103955.
66. S. Ahmad, Y. Liao, A. Hussain, Q. Zhang, E.-X. Ding, H. Jiang and E. I. Kauppinen, *Carbon*, 2019, 149, 318–327.
67. D. V. Krasnikov, B. Y. Zabelich, V. Y. Iakovlev, P. Tsapenko, S. A. Romanov, A. A. Alekseeva, A. K. Grebenko and A. G. Nasibulin, *Chem. Eng. J.*, 2019, 372, 462–470.
68. C. G. Granqvist and R. A. Buhrman, *J. Appl. Phys.*, 1976, 47, 2200–2219.
69. C. Hoecker, F. Smail, M. Bajada, M. Pick and A. Boies, *Carbon*, 2016, 96, 116–124.
70. M. Vila, S. Hong, S. Park, A. Mikhalchan, B.-C. Ku, J. Y. Hwang and J. J. Vilatela, *ACS Appl. Nano Mater.*, 2021, 4, 6947–6955.
71. B. Alemán and J. J. Vilatela, *Carbon*, 2019, 149, 512–518.
72. Kaniyoor, J. Bulmer, T. Gspann, J. Mizen, J. Ryley, P. Kiley, J. Terrones, C. Miranda-Reyes, G. Divitini, M. Sparkes, B. O'Neill, A. Windle and J. A. Elliott, *Nanoscale*, 2019, 11, 18483–18495.
73. M. Mujica, G. Tutuncuoglu, P. P. Shetty, A. T. Mohabir, E. V. Woods, V. Breedveld, S. H. Behrens and M. A. Filler, ACS Appl. Nano Mater., 2020, 3, 905–913.
74. G. G. Tibbetts, in Carbon Filaments and Nanotubes: Common Origins, Differing Applications?, ed. L. P. Biró, C. A. Bernardo, G. G. Tibbetts and P. Lambin, Springer Netherlands, Dordrecht, 2001, pp. 1–9.
75. J.-P. Lange, *Catal. Sci. Technol.*, 2016, 6, 4759–4767.
76. Javadi, M. Soltanieh, S. Sahebdelfar, D. Bastani and K. Javadi, ASME International Mechanical Engineering Congress and Exposition, 2006, vol. 47772, pp. 69–77.
77. Boje, J. Akroyd, S. Sutcliffe and M. Kraft, *Chem. Eng. Sci.*, 2020, 219, 115615.
78. J. Feng, G. Biskos and A. Schmidt-Ott, *Sci. Rep.*, 2015, 5, 1–9.
79. M. Boies, C. Hoecker, A. Bhalerao, N. Kateris, J. de La Verpilliere, B. Graves and F. Smail, *Small*, 2019, 15, 1900520.
80. M. Li, G. W. Mulholland and M. R. Zachariah, *Aerosol Sci. Technol.*, 2014, 48, 139–141.
81. N. Kateris, P. Kloza, R. Qiao, J. A. Elliott and A. M. Boies, *J. Phys. Chem. C*, 2020, 124, 8359–8370.
82. K. Mustonen, PhD thesis, Department of Applied Physics, Aalto University School of Science, 2015.
83. D. E. Tsentalovich, R. J. Headrick, F. Mirri, J. Hao, N. Behabtu, C. C. Young and M. Pasquali, *ACS Appl. Mater. Interfaces*, 2017, 9, 36189–36198.
84. N. Fakhri, D. A. Tsyboulski, L. Cognet, R. B. Weisman and M. Pasquali, *Proc. Natl. Acad. Sci. U. S. A.*, 2009, 106, 14219–14223.
85. K. Chakraborty, T. Plyhm, M. Barbezat, A. Necola and G. P. Terrasi, *J. Nanopart. Res.*, 2011, 13, 6493–6506.
86. F. Smail, A. Boies and A. Windle, *Carbon*, 2019, 152, 218–232.
87. J. J. Vilatela and A. H. Windle, *Adv. Mater.*, 2010, 22, 4959–4963.
88. V. Negi and R. C. Picu, *Soft Matter*, 2021, 17, 10186–10197.
89. Mikhalchan, C. Madrona, L. Arévalo, M. Malfois and J. J. Vilatela, *Carbon*, 2022, 197, 368–377.
90. S. Borodulina, A. Kulachenko, M. Nygårds and S. Galland, Nord. Pulp Pap. Res. J., 2012, 27, 318–328.
91. T. F. Cooke, *J. Am. Ceram. Soc.*, 1991, 74, 2959–2978.
92. Q. Zhang, N. Wei, P. Laiho and E. I. Kauppinen, *Top. Curr. Chem.*, 2017, 375, 90.
93. O. T. Zaremba, A. E. Goldt, E. M. Khabushev, A. S. Anisimov and A. G. Nasibulin, *Mater. Sci. Eng., B*, 2022, 278, 115648.
94. P. Hou, F. Zhang, L. Zhang, C. Liu and H. Cheng, *Adv. Funct. Mater.*, 2022, 32, 2108541.
95. S. De and J. N. Coleman, *MRS Bull.*, 2011, 36, 774–781.
96. J. Di, X. Zhang, Z. Yong, Y. Zhang, D. Li, R. Li and Q. Li, *Adv. Mater.*, 2016, 28, 10529–10538.

97. X. Chen, I. Ali, L. Song, P. Song, Y. Zhang, S. Maria, S. Nazmus, W. Yang, H. N. Dhakal, H. Li, M. Sain and S. Ramakrishna, Renewable Sustainable Energy Rev., 2020, 134, 110085.
98. S.-H. Park, P. J. King, R. Tian, C. S. Boland, J. Coelho, C. Zhang, P. McBean, N. McEvoy, M. P. Kremer, D. Daly, J. N. Coleman and V. Nicolosi, Nat. Energy, 2019, 4, 560–567.
99. R. Tian, N. Alcala, S. J. K. O'Neill, D. V. Horvath, J. Coelho, A. J. Griffin, Y. Zhang, V. Nicolosi, C. O'Dwyer and J. N. Coleman, ACS Appl. Energy Mater., 2020, 3, 2966–2974.
100. C. Jo, A. S. Groombridge, J. De La Verpilliere, J. T. Lee, Y. Son, H.-L. Liang, A. M. Boies and M. De Volder, ACS Nano, 2020, 14, 698–707.
101. M. Rana, C. Santos, A. Monreal-Bernal and J. J. Vilatela, Synthesis and Applications of Nanocarbons, Wiley, 2020, pp.149–200.
102. M. Rana, V. Sai Avvaru, N. Boaretto, V. A. de la Peña O'Shea, R. Marcilla, V. Etacheri and J. J. Vilatela, J. Mater. Chem. A, 2019, 7, 26596–26606.
103. S. Zeng, H. Chen, H. Wang, X. Tong, M. Chen, J. Di and Q. Li, Small, 2017, 13, 1–8.
104. K. Evanoff, J. Benson, M. Schauer, I. Kovalenko, D. Lashmore, W. J. Ready and G. Yushin, ACS Nano, 2012, 6, 9837–9845.
105. A. Moya, N. Kemnade, M. R. Osorio, A. Cherevan, D. Granados, D. Eder and J. J. Vilatela, J. Mater. Chem. A, 2017, 5, 24695–24706.
106. A. Martínez-Muíño, M. Rana, J. J. Vilatela and R. D. Costa, Nanoscale Adv., 2020, 2, 4400–4409.
107. A. Pendashteh, J. Palma, M. Anderson, J. J. Vilatela and R. Marcilla, ACS Appl. Energy Mater., 2018, 1, 2434–2439.
108. H. Liu, J. Zhou, C. Wu, C. Wang, Y. Zhang, D. Liu, Y. Lin, H. Jiang and L. Song, ACS Sustainable Chem. Eng., 2018, 6, 2911–2915.
109. Z. Xu, Z. Zhou, B. Li, G. Wang and P. W. Leu, J. Phys. Chem. C, 2020, 124, 8689–8696.
110. M. Rana, N. Boaretto, A. Mikhalchan, M. Vila Santos, R. Marcilla and J. J. Vilatela, ACS Appl. Energy Mater., 2021, 4, 5668–5676.
111. Z. Pan, X. Liu, J. Yang, X. Li, Z. Liu, X. J. Loh and J. Wang, Adv. Energy Mater., 2021, 11, 2100608.
112. S. Upama, A. Mikhalchan, L. Arevalo, M. Rana, A. Pendashteh, M. J. Green and J. J. Vilatela, ACS Appl. Mater. Interfaces, 2023, 15, 5590–5599.
113. N. Boaretto, J. Almenara, A. Mikhalchan, R. Marcilla and J. J. Vilatela, ACS Appl. Energy Mater., 2019, 2, 5889–5899.
114. E. Senokos, M. Rana, M. Vila, J. Fernandez-Cestau, R. D. Costa, R. Marcilla and J. J. Vilatela, Nanoscale, 2020,
115. C. Santos, I. V. Rodríguez, J. J. Lado, M. Vila, E. García-Quismondo, M. A. Anderson, J. Palma and J. J. Vilatela, Carbon, 2021, 176, 390–399.
116. A. Monreal-Bernal, J. J. Vilatela and R. D. Costa, Carbon, 2019, 141, 488–496.
117. A. Monreal-Bernal and J. J. Vilatela, ChemPlusChem, 2018, 83, 285–293.
118. Z. Niu, H. Dong, B. Zhu, J. Li, H. H. Hng, W. Zhou, X. Chen and S. Xie, Adv. Mater., 2013, 25, 1058–1064.
119. M. Rana, Y. Ou, C. Meng, F. Sket, C. González and J. J. Vilatela, Multifunct. Mater., 2020, 3, 015001.
120. J. Di, Z. Yong, X. Zheng, B. Sun and Q. Li, Small, 2013, 9, 1367–1372.
121. X. Ren, K. Turcheniuk, D. Lewis, W. Fu, A. Magasinski, M. W. Schauer and G. Yushin, Small, 2018, 14, 1–9.
122. R. Li, X. Xiang, X. Tong, J. Zou and Q. Li, Adv. Mater., 2015, 27, 3831–3835.
123. Z. Zhang, Z. Yang, Z. Wu, G. Guan, S. Pan, Y. Zhang, H. Li, J. Deng, B. Sun and H. Peng, Adv. Energy Mater., 2014, 4, 1–6.
124. J. Ren, Y. Zhang, W. Bai, X. Chen, Z. Zhang, X. Fang, W. Weng, Y. Wang and H. Peng, Angew. Chem., Int. Ed., 2014, 53, 7864–7869.
125. X. Chen, L. Qiu, J. Ren, G. Guan, H. Lin, Z. Zhang, P. Chen, Y. Wang and H. Peng, Adv. Mater., 2013, 25, 6436–6441.
126. G. Zang, J. Zhang, S. Xu and Y. Xing, Energy, 2021, 218, 119504.
127. F. Wu, J. Maier and Y. Yu, Chem. Soc. Rev., 2020, 49, 1569–1614.
128. Y. Liu, R. Zhang, J. Wang and Y. Wang, iScience, 2021, 24, 102332.

129. A. N. L. Energy Systems Division, Material and Energy Flows in the Materials Production, Assembly, and End-of- Life Stages of the Automotive Lithium-Ion Battery Life Cycle, 2012.
130. A. Yella, S. Mathew, S. Aghazada, P. Comte, M. Gratzel and M. K. Nazeeruddin, J. Mater. Chem. C, 2017, 5, 2833– 2843.
131. N. P. Brandon and D. J. Brett, Philos. Trans. R. Soc., A, 2006, 364, 147–159.
132. A. Mikhalchan and J. J. Vilatela, Carbon, 2019, 150, 191– 215.
133. B. Natarajan, Compos. Sci. Technol., 2022, 225, 109501.
134. S. Banerjee, J. Luginsland and P. Zhang, 2018 IEEE International Conference on Plasma Science (ICOPS), 2018, p. 1.
135. A. Lekawa-Raus, J. Patmore, L. Kurzepa, J. Bulmer and K. Koziol, Adv. Funct. Mater., 2014, 24, 3661–3682.
136. J. Zou, X. Zhang, C. Xu, J. Zhao, Y. T. Zhu and Q. Li, Carbon, 2017, 121, 242–247.